\documentclass[usenatbib]{mnras}
\usepackage{graphicx}
\usepackage{txfonts}

\usepackage{url}
\usepackage{graphicx}
\usepackage[hyphenbreaks]{breakurl}
\usepackage{float}
\usepackage{subfig}
\usepackage{multirow}
\usepackage{natbib}
\usepackage{xcolor}
\usepackage{soul}

\title{Implications of spicule activity on coronal loop heating and catastrophic cooling}

\author[]
{V.N. Nived,$^{1,2}$ E. Scullion,$^{3}$ J. G. Doyle,$^{1}$ R. Susino,$^{4}$ P. Antolin,$^{3}$ D. Spadaro,$^{5}$  
\newauthor C. Sasso,$^{6}$ S. Sahin,$^{3}$ M. Mathioudakis,$^{2}$ \\
$^{1}$Armagh Observatory \& Planetarium, College Hill, Armagh, BT61 9DG, N. Ireland\\
$^{2}$Astrophysics Research Centre, School of Mathematics and Physics, Queens University, Belfast, BT7 1NN, N. Ireland\\
$^{3}$ Department of Mathematics, Physics and Electrical Engineering, Northumbria University, Newcastle Upon Tyne NE1 8ST, UK\\
$^{4}$ Istituto Nazionale di Astroﬁsica (INAF), Osservatorio Astroﬁsico di Torino, Via Osservatorio, 20,10125, Pino Torinese (TO), Italy\\
$^{5}$INAF–Turin Astrophysical Observatory, Via Osservatorio 20, 10025 Pino Torinese (TO), Italy\\
$^{6}$INAF–Osservatorio Astronomico di Capodimonte, Salita Moiariello 16, I–80131 Napoli, Italy\\}

\begin{document}
\outer\def\gtae {$\buildrel {\lower3pt\hbox{$>$}} \over 
{\lower2pt\hbox{$\sim$}} $}
\outer\def\ltae {$\buildrel {\lower3pt\hbox{$<$}} \over 
{\lower2pt\hbox{$\sim$}} $}
\newcommand{\Msun}{$M_{\odot}$}
\newcommand{\lsun}{$L_{\odot}$}
\newcommand{\Rsun}{$R_{\odot}$}
\newcommand{\solar}{${\odot}$}
\newcommand{\kep}{\sl Kepler}
\newcommand{\ktwo}{\sl K2}
\newcommand{\tess}{\sl TESS}
\newcommand{\swift}{\it Swift}
\newcommand{\Porb}{P_{\rm orb}}
\newcommand{\nuorb}{\nu_{\rm orb}}
\newcommand{\eplus}{\epsilon_+}
\newcommand{\eminus}{\epsilon_-}
\newcommand{\cd}{{\rm\ c\ d^{-1}}}
\newcommand{\MdotL}{\dot M_{\rm L1}}
\newcommand{\Mdot}{$\dot M$}
\newcommand{\Mdotsolar}{\dot{M_{\odot}} yr$^{-1}$}
\newcommand{\Ldisk}{L_{\rm disk}}
\newcommand{\src}{KIC 9202990}
\newcommand{\ergscm} {erg s$^{-1}$ cm$^{-2}$}
\newcommand{\rchi}{$\chi^{2}_{\nu}$}
\newcommand{\chisq}{$\chi^{2}$}
\newcommand{\pcmsq} {cm$^{-2}$}

\newcommand{\patrick}[1]{\textcolor{orange}{#1}}

\providecommand{\lum}{\ensuremath{{\cal L}}}
\providecommand{\mg}{\ensuremath{M_{\rm G}}}
\providecommand{\bcg}{\ensuremath{BC_{\rm G}}}
\providecommand{\mbolsun}{\ensuremath{M_{{\rm bol}{\odot}}}}
\providecommand{\teff}{\ensuremath{T_{\rm eff}}}

\maketitle
\begin{abstract}

We report on the properties of coronal loop foot-point heating with observations at the highest resolution, from the CRisp Imaging Spectro-Polarimeter (CRISP) located at the Swedish 1-m Solar Telescope (SST) and co-aligned NASA Solar Dynamics Observatory (SDO) observations, of Type II spicules in the chromosphere and their signatures in the EUV corona. Here, we address one important issue, as to why there is not always a one-to-one correspondence, between Type II spicules and hot coronal plasma signatures, i.e. beyond TR temperatures. 
We do not detect any difference in their spectral properties in a quiet Sun region compared to a region dominated by coronal loops. On the other hand, the number density close to the foot-points in the active region is found to be an order of magnitude higher than in the quiet Sun case. A differential emission measure analysis reveals a peak at $\sim 5 \times 10^5$ K on the order of 
10$^{22}$~cm$^{-5}$~K$^{-1}$. Using this result as a constraint, we conduct numerical simulations and show that with an energy input of $1.25 \times 10^{24}$ erg (corresponding to $\sim$10 RBEs contributing to the burst) we manage to reproduce the observation very closely.
However, simulation runs with lower thermal energy input do not reproduce the synthetic AIA 171~\AA\ signatures, indicating that there is a critical number of spicules required in order to account for the AIA 171~\AA\ signatures in the simulation.
Furthermore, the higher energy ($1.25 \times 10^{24}$ ergs) simulations reproduce catastrophic cooling with a cycle duration of $\sim$5 hours, matching a periodicity we observe in the EUV observations.  
\end{abstract}


\section{Introduction}

The prevailing theories for the heating of coronal active regions include magnetohydrodynamic waves and magnetic reconnection via nano-flaring or braiding motions in the foot-points of flux tubes, plus mass loading via spicule activity \citep{Srivastava_2017, Nived2020,Cirtain2013, Mart2018}. Type II spicules move faster (50 to 150 km s$^{-1}$) and have shorter lifetimes (10 to 150 s) compared to Type I spicules \citep{type_2}. Unlike Type I spicules they disappear without evidence for down-falling material as seen in H$\alpha$. However, they do show a falling phase in observations taken with IRIS and AIA filters \citep{Pereira_2014,skogsrud_2015}. Furthermore, they are also considered as a potential candidate for coronal heating. Various MHD wave modes are also observed in spicules and can contribute to the heating of the corona. \cite{bart_2007} suggested that the transverse oscillation of a spicule with respect to its axis is associated with Alfv\'en waves. \cite{shetye_2016} reported the presence of torsional motions in spicules while \cite{Srivastava_2017} suggested that these motions can be interpreted as torsional Alfv\'en waves that are associated with high-frequency drivers transferring energy into the corona. This energy input is sufficient to heat the corona to million degrees. However, the processes by which these waves are generated and dissipated are unclear. \citet{Antolin_2018ApJ...856...44A} showed that the observed motions could also be explained by kink waves that trigger the Kelvin-Helmholtz instability (KHI). In this case, the KHI allows the dissipation of the wave energy through the generated turbulence. Observations with high spatio-temporal resolution are therefore required to understand the properties of spicules in great detail and discern the heating mechanisms. \cite{van2015} looked at rapid blue- or red-shifted excursions (RBEs or RREs) and suggested that at least part of the diffuse halo around network regions can be attributed to Type II spicules. On the other hand, \cite{Klimchuk2012} suggested that only a small fraction of the hot plasma can be supplied by spicules.

The spicule thermal energy is often calculated assuming a Type II mass density of $3\times 10^{-10}$ kg m$^{-3}$, a volume based on a cylindrical geometry with a radius of 100 km and the spicule's chromospheric length as 10000 km. Considering that the chromosphere is hydrogen dominated (90\%), the number of electrons is approximately equal to the number of protons. Taking the spicule proton mass, and an upper chromosphere temperature of T $\sim$15,000 K, a thermal energy of $1.8 \times 10^{23}$ erg can be derived. Assuming a spicule velocity of 50 km s$^{-1}$, the derived kinetic energy is $1.2 \times 10^{24}$ ergs. 
Others have derived different energy estimates, e.g. \cite{bart_2009} derived an energy of $10^{23}$ erg for Type II spicules, while \cite{Klimchuk2012} derived $8 \times 10^{22}$ erg based on an emission measure distribution. \citet{Mamedov2016} discussed several possibilities for energy input due to both classical and Type II spicules, which in some instances are an order of magnitude larger than the above values. In a recent work, \citet{Tanmoy2019} performed an order-of-magnitude estimate of the kinetic energy of spicules ($1.2 \times 10^{23}$ erg), and the maximum available magnetic energy ($2.5 \times 10^{25}$ erg) at the foot-points of spicules based on H$\alpha$ spectral scans and magnetic field data from the Goode Solar Telescope.

Most of these are upper limits as it assumes 100\% of the energy is converted to heat; using a lower limit of 10–20\% energy conversion gives between $10^{22}$ erg and a maximum of $10^{23}$ erg for the spicule thermal energy.

Type II spicules are omnipresent and carry a large amount of magnetic energy \citep{Liu2019}. Some are found to have on-disk counterparts and are observed as rapid blue-shifted excursions (RBEs). \citet{Sekse_2013} used a high cadence data-set and found that the RBE lifetime ranges from 5 to 60 s and their transverse velocities may reach up to 55 km s$^{-1}$. Data from the Interface Region Imaging Spectrograph indicate increased line broadening, suggesting that they are heated to at least transition region temperatures \citep{Henr2016}. The formation process of Type II spicules is thought to affect the upper atmosphere via flows, waves and shocks. In some instances, they are related to red-blue line asymmetries in UV data, and/or to propagating coronal disturbances \citep{Mart2018}. \citet{Yur2013} reported that the occurrence of packets of Type II spicules is generally correlated with the appearance of new, mixed or unipolar fields in close proximity to network fields. \citet{Tanmoy2019} observed spicules emerging within minutes of the appearance of opposite-polarity magnetic flux around dominant-polarity magnetic field concentrations showing heating of the adjacent corona.

In section 2, we present observations of RBEs from the Swedish Solar Telescope (SST) coupled with simultaneous data from the Solar Dynamic Observatory (SDO). The observed region included both loop foot-points and a region devoid of warm loops. In Section 3, we discuss the detection and tracing of the RBEs, while the results are outlined in Section 4. In Section 5, we outline simulations of impulsive heating in a coronal loop. Section 6 discusses the simulated coronal condensation and implications for coronal heating. 
\begin{figure}
\centering
\vspace*{-1.5cm}
\includegraphics[width=0.5\textwidth]{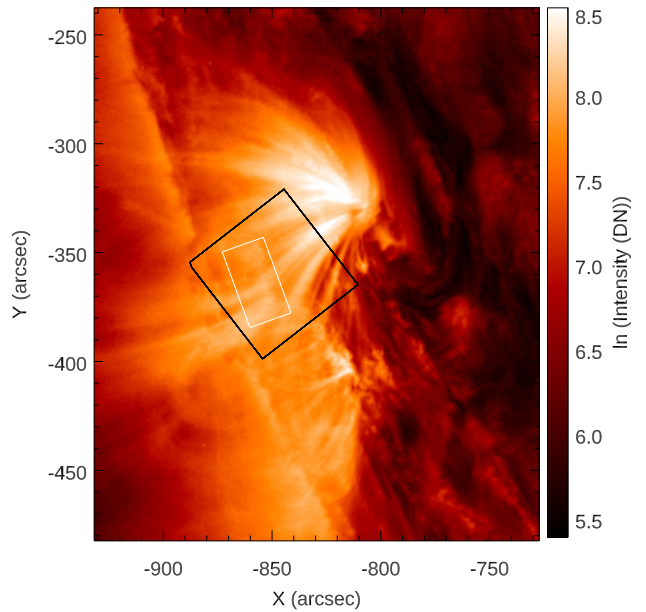}
\vspace*{-1.0cm}

\caption{AIA 171 {\AA} EUV channel image of the target region. The black box represents the FOV of the SST observation. The white box bounds the region of interest, which includes coronal loops and quiet Sun.} 
\label{fig0}
\end{figure}


\begin{figure}
\centering
\vspace*{-1cm}
\hspace*{-1.5cm}
\includegraphics[width=0.65\textwidth]{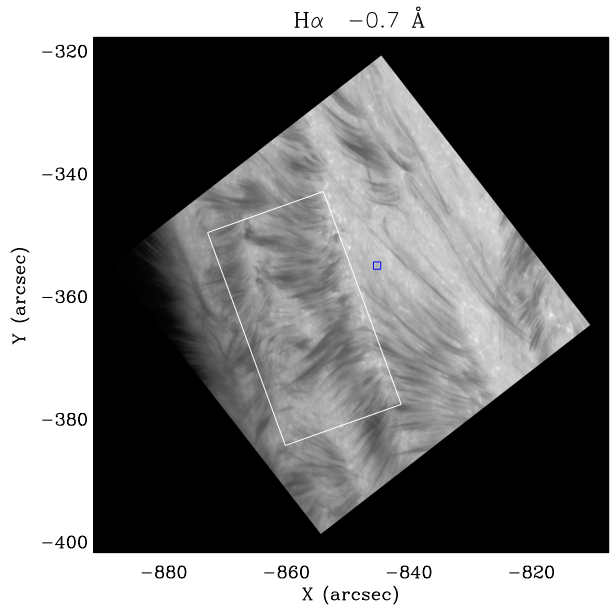}
\vspace*{-1cm}

\caption{CRISP observation of H$\alpha$ at 6562.3 {\AA} ($-$0.7 {\AA} from line centre). The white box represents the FOV selected for detecting the RBEs (region of interest). The small blue box represents the region used for obtaining the reference H$\alpha$ profile}

\label{fig1}
\end{figure}

\section{Observations \& data reduction}

CRISP is an imaging spectro-polarimeter that includes a dual Fabry-Pérot interferometer (FPI) \citep{Scharmer_2006}. During the observations, CRISP was set to cycle continuously through 41 line positions from 6561.28 \AA\ ($-$1.72 {\AA} from line centre) in the H$\alpha$ blue wing through line centre to 6564.72 \AA\ ($+$1.72 {\AA} from line centre) in the red wing, in equidistant steps of 85 m\AA. We used 8 exposures per wavelength position. For H$\alpha$ 6563 \AA, the transmission FWHM of CRISP is 66 m\AA\ and the pre-filter is 4.9 \AA. By applying the Multi-Object Multi-Frame Blind Deconvolution (MOMFBD: \citet{Noort2005} image restoration technique, {the time series images are corrected for residual atmospheric distortions that were not corrected by the adaptive optics system.
 We refer to \cite{van_2008}, \cite{lin_2012}, \cite{Sekse_2013} and \cite{Scullion2015}, for more details on the post MOMFBD processing applied to the data under investigation. The fully processed data were analysed further with CRISPEX, a code for analysis of multi-dimensional data-cubes \citep{vissers_2012}. The pixel size of the H$\alpha$ images is 0.0592 arcsec, providing approximately 10 times more resolving power than that of the SDO/AIA images (0.6 arcsec per pixel). 
 \begin{figure}
  \centering
\hspace*{-1cm}
\includegraphics[width=0.5\textwidth]{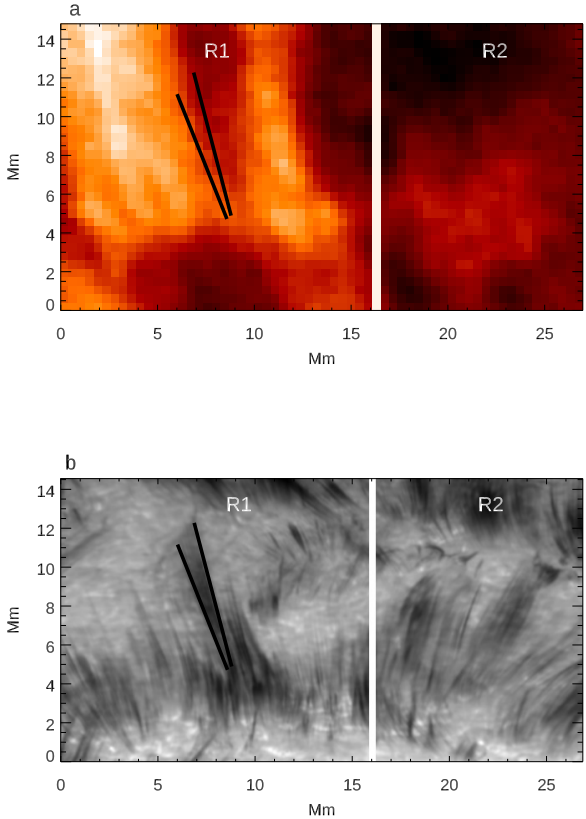}
\caption{Panel (a) : AIA 171 {\AA} image of the FOV selected for the detection of RBEs. Panel (b): Corresponding H$\alpha$ blue wing image. The white line separate the region that contain the coronal loop foot-points (R1) and the region that does not contain any loops (R2). The pixel scale of the SST is a factor of ten better than AIA. The black lines mark the location of large spicule in panels (a) and (b).}
\label{fig2}
\end{figure}


The standard SolarSoft reduction pipeline is applied to the raw SDO/AIA images \citep{Lemen2012}. SDO/AIA observes the Sun in eight different temperature channels with an uninterrupted cadence of 12 s. The AIA standard reduction uses the following stages, namely, decompression and reversion, dead-time correction, local gain correction, flat-field correction, cosmic ray removal and hot pixel correction. To achieve sub-AIA pixel accuracy in the co-alignment of AIA with CRISP narrow-band images we cross-correlate photospheric bright points in the simultaneous CRISP wide-band images and AIA continuum ($\sim$5000 K) and EUV 171 \AA\ coincident images.

An overview of the Field-of-View (FOV) of the CRISP observation is presented in Figure \ref{fig0}. The observations\footnote{ The data underlying this article were provided by the solar physics group at RoCS (University of Oslo) with different aspects reported by \cite{Scullion2015}.}
were carried out very close to the limb in the emerging active region NOAA 11084.  In Figure \ref{fig1}, we show the H$\alpha$ blue wing image of the target region observed by CRISP on 27 June 11:01 UT. The CRISP observation started at 10:58:54 UT on 27 June 2010 and continued till 11:30:01 UT. During this period CRISP recorded 208 images of the target region in each line position with a cadence of $\sim $9.02 s.

The white box overlaid on Figure \ref{fig1} bounds the region of interest in this study, which contains dark spicular structures protruding outwards from the solar surface. In Figure \ref{fig2}, we present the detailed view of the white box plotted in Figure \ref{fig1} and the corresponding AIA 171~\AA\ image. Co-aligned AIA 171~\AA\ images of the region of interest can be segmented into two distinct regions based on the intensity, namely R1 and R2 as labelled in Figure ~\ref{fig2}. The corresponding AIA 171~\AA\ images reveal that region R1 contains coronal loop foot-points while region R2 does not for over a 20 hr interval (we should note that R2 does contain coronal emission, although not as intense as in R1). This particular FOV provides us with an excellent opportunity to study whether RBEs affect the coronal loops. This region is cropped and investigated in great detail in the remaining sections.

From the H$\alpha$ blue wing images, it is clear that regions R1 and R2 both contain RBEs. RBEs appear as a broadening in the blue wing of H$\alpha$ with an unaffected red wing, therefore they appear as a dark structure in the blue wing images. Hydrogen lines are very sensitive to thermal broadening because of the lower atomic mass. The thermal broadening can also appear as dark absorbing features in the H$\alpha$ blue wing and red wing images. Therefore one must be careful while detecting RBEs or RREs. In order to detect the RBEs in these observations, we developed an automated algorithm based on the residual method. The details of the algorithm are provided in the next section.

  \begin{figure}
  \centering
   \hspace*{-1cm}
\includegraphics[width=0.55\textwidth]{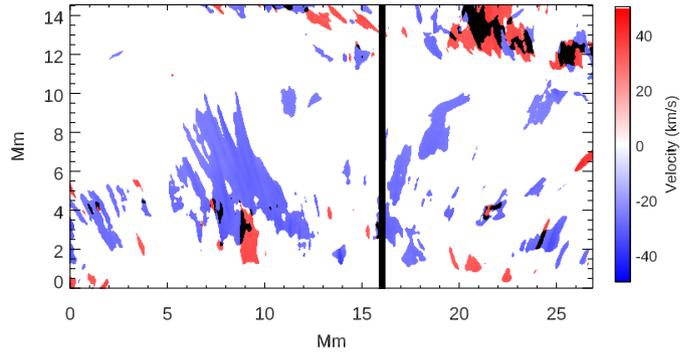}
\caption{Doppler map of RBEs and RREs. The black line separated the region R1 and R2. The blue region in the velocity map represents the RBEs while the red region represents the RREs. The black pixels shows the location of strong broadening.} 
\label{fig4}
\end{figure}



  \begin{figure}
  \centering
  \hspace*{-1cm}
\includegraphics[width=0.55\textwidth]{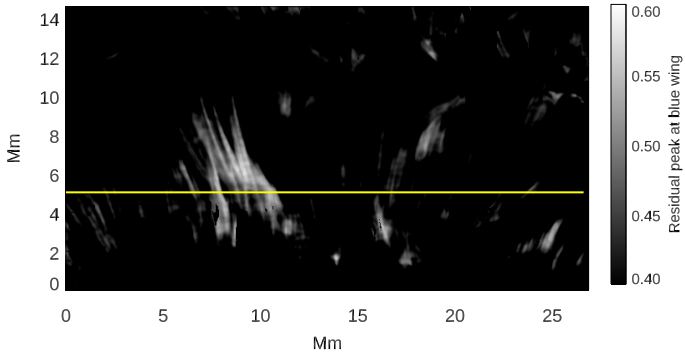}
\caption{Map of peak value of residual in the blue wing. Bright regions represent the location of RBEs. The yellow line depicts the horizontal slit chosen for the tracking of RBEs.} 
\label{fig5}
\end{figure}



  \begin{figure}
  \centering
    \hspace*{-1cm}
\includegraphics[width=0.55\textwidth]{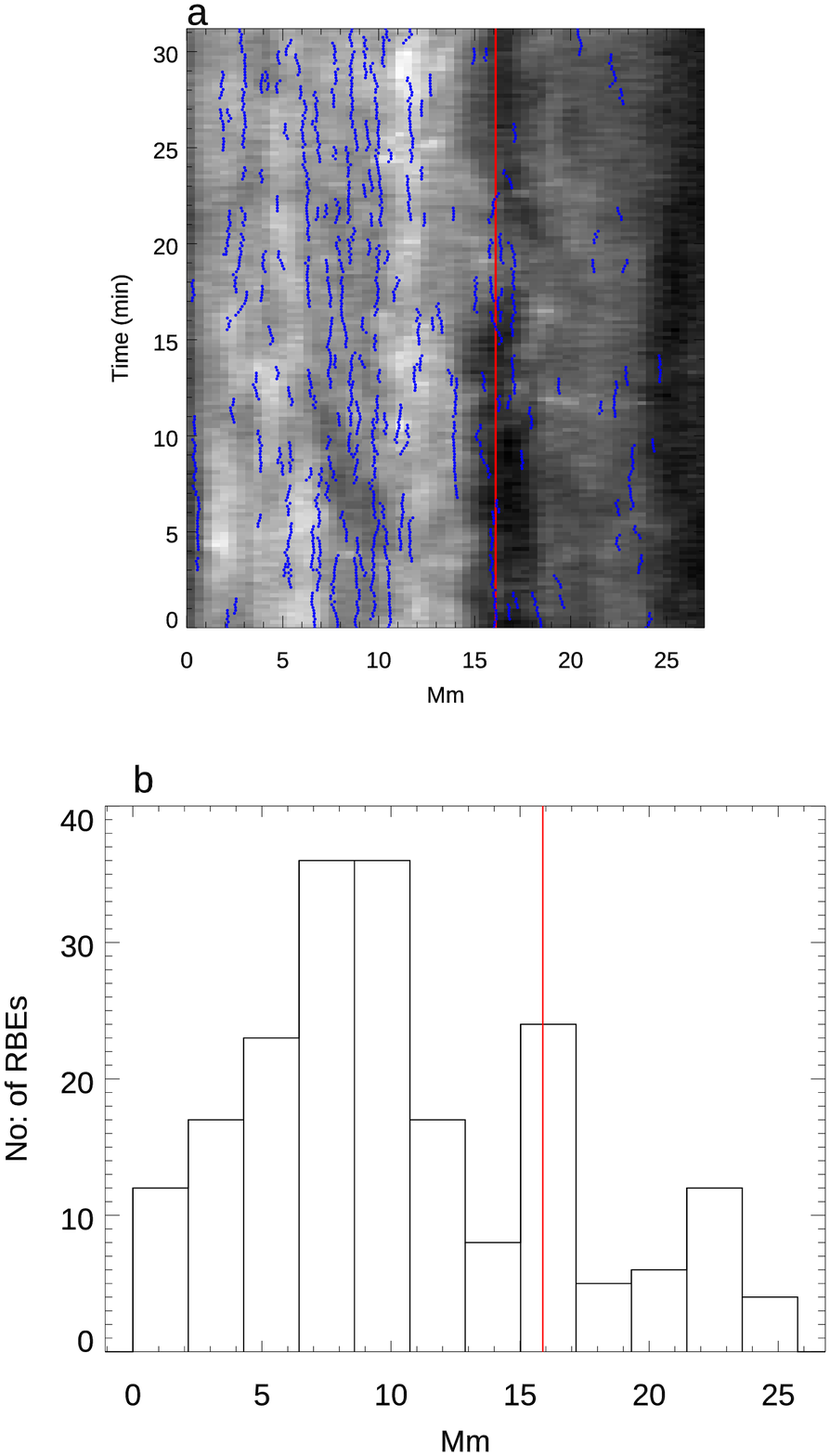}
\caption{Statistics of RBEs in region R1 and R2. panel (a): Time-distance plot of the RBEs (blue dots) detected along the horizontal slit at y-location $\sim$ 5 Mm overlaid on AIA 171~\AA\ time-distance map.  Panel (b): Histogram for the number of RBEs as a function of horizontal location. The red line in both panels separates regions R1 from R2. There is a higher number of RBEs in region R1 compared to R2.}
\label{fig6}
\end{figure}



\section{RBE detection \& tracking}

\subsection{RBE detection}\label{detection}

We have undertaken the following steps for the detection of RBEs:
\begin{enumerate}

\item The spectra are normalised with respect to the continuum intensity. A reference profile for the H$\alpha$ line is obtained by averaging the spectrum in a quiet Sun region, see the blue box in Figure \ref{fig1}.

\item Once the reference profile is calculated, the code scans through the normalised H$\alpha$ profile of each pixel in the selected FOV and subtracts the reference profile from it. This gives a residual profile for each pixel in the FOV. In most cases, the residual profile will have two peaks; one in the blue wing and the other in the red wing due to the effects of line broadening. 

\item The code then calculates separately in the blue and red wing the peak value and the wavelength corresponding to the peak by fitting a single Gaussian to the residual profile. By manual inspection, we defined the RBEs as the pixels with a residual peak above 0.4 (corresponding to 13$\sigma$) in the blue wing. Similarly, the RREs are the pixels with a residual peak above 0.4 in the red wing. If we have peaks that are above 0.4 in both the wings, the corresponding pixel is considered as due to broadening. We also tried values lower than 0.4, but this did not affect the number of RBEs in R1 compared to R2.

\end{enumerate}


  \begin{figure}
    \hspace*{-0.5cm}
\includegraphics[width=0.52\textwidth]{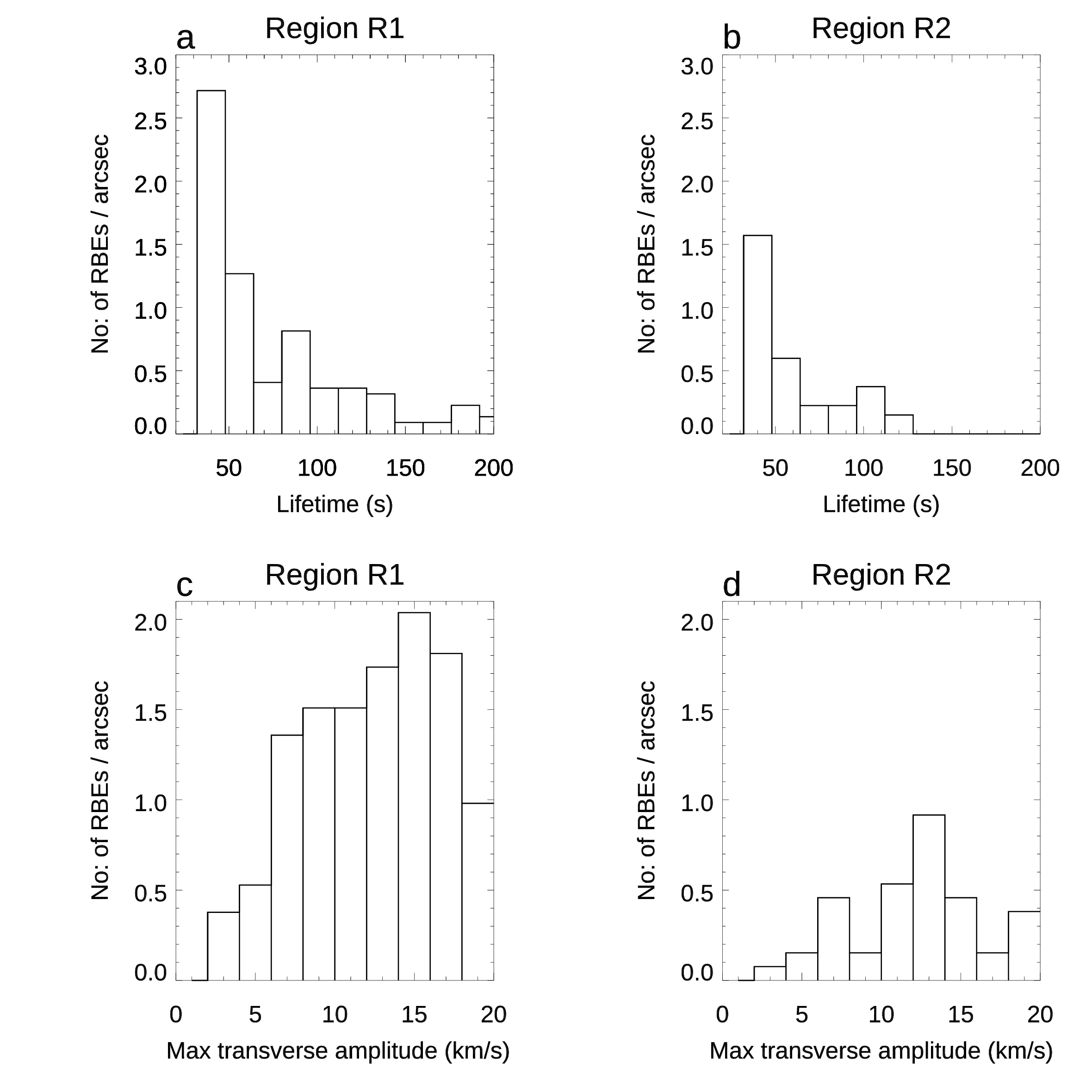}
\caption{Panel (a): Distribution of lifetime of the RBEs in the loop foot-point region (R1) in the AIA 171 \AA\ channel. Panel (b): Distribution of lifetime of the RBEs in the quieter region (R2) in the AIA 171 \AA\ channel. Panel (c): Distribution of maximum transverse velocities in the plane-of-the-sky of the RBEs detected in the loop foot-point region (R1). Panel (d):  Distribution of maximum transverse velocities of the RBEs in the less intense region (R2).} 
\label{fig55}
\end{figure}


In Figure \ref{fig4}, we plot the Doppler velocity map for the selected region for the first frame in the time series. The Doppler shift was found by subtracting the line core wavelength from the wavelength corresponding to the residual peak. Since our observations were taken very close to the limb, the differential rotation of the Sun can affect the velocity measurement. In order to correct for this effect, we calculated the speed of rotation at the latitude corresponding to the observation and subtracted this velocity from the observed Doppler velocity. The blue region in the velocity map represents the RBEs while the red region represents the RREs. The white pixels in the map is where the peak of the residual is less than 0.4 in both wings. The black pixels show the location of strong broadening. The Doppler map reveals that there are many more RBEs in region R1 than in R2 (this is better seen in later figures, e.g.  Figure~\ref{fig6}). The next step is to obtain and compare the number of RBEs in region R1 and R2.

\subsection{Tracking RBEs}\label{tracking}
The main aim of this paper is to study the statistics of RBEs in regions R1 and R2. In order to do this, we need to track the location of individual RBEs and their evolution as a function of time. For this analysis, we used the peak residual map which is created from the peak of the residual profile in the blue wing. An example of a peak residual map is shown in Figure \ref{fig5}. The bright region is where the broadening in the blue wing is stronger. Therefore the local enhancement in the brightness can be considered as the location of the RBEs. The tracking of the RBEs is carried out along a horizontal slit at y-location $\sim$ 5 Mm as shown in Figure \ref{fig5}. This location corresponds to the location of the highest number of RBEs detected along the horizontal slit. The main steps of the tracking algorithm is explained below.

\begin{enumerate}

\item The tracking algorithm searches through the horizontal slit for the local enhancement in the peak residual value. A local enhancement is defined as the region with a continuous increase in residual peak in at least 5 neighbouring pixels, followed by a continuous reduction in residual peak in at least 5 neighbouring pixels; i.e. we select the location of the brightest point as the location of the RBE. 
\item Once the local maximum is identified, the code fits a Gaussian profile to the local maximum and calculates the position of the RBEs in sub spatial resolution. We repeat this process for each time-step and track the evolution of the RBEs in space and time. 
\item Since the RBEs undergo sideways (horizontal) motions, the location of the RBEs may change significantly from one time step to the next. We imposed a condition that the horizontal shift should be less than 4 pixels in two consecutive frames. This means that if the difference in the horizontal position is less than 4 pixels in two consecutive frames corresponding to a shift of up to 20 km s$^{-1}$, the code assumes that the RBE seen in the next frame is the same RBE as seen in the previous frame. 
We checked using only 2 pixels which equates to a shift of only 10 km s$^{-1}$. Such a value would mean that we miss the high-velocity events. The disadvantage is that we fail to separate nearby RBEs. Overall, 4 pixels is a better choice. 
\item Finally, we applied a condition that the minimum lifetime of RBEs should be at least three time-steps. After applying all these conditions, a time-distance plot for RBEs is obtained. The time-distance plot can be used to calculate the horizontal movement as well as the lifetime of the RBEs. 

\end{enumerate}

 \section{Results}
 
\subsection{Number Density}\label{number} 

\indent The results obtained by following the RBE tracking scheme are shown in Figure~\ref{fig6}. In panel (a), we overplot the time-distance plot for RBEs on an AIA 171~\AA\ time-distance map obtained from the same location. The blue dots in panel (a) represents the RBEs detected along the horizontal slit at $y_{pix} = 115$ (see Section \ref{tracking} above). The red line represents the boundary between the quiet Sun and the loop foot-point region in the AIA 171~\AA\ channel. From the time-distance plot, it is clear that the number of RBEs in the bright region is larger than in the region of lower brightness. This is further confirmed by plotting the histograms representing the number density of events as shown in panel (b). The bin size of the histogram is $\sim 2.1$ Mm.  Our analysis reveals that the loop foot-point region R1 contains on average a factor of 3 more RBEs than in the quiet Sun region (see Table 1). However, region R1 also contains quiet Sun plasma. Looking at panel (b) in Figure~\ref{fig6}, we see some locations in R1 which has a factor of 7--9 more RBEs than in region R2. 

  \begin{figure}
      \hspace*{-0.5cm}
\includegraphics[width=0.52\textwidth]{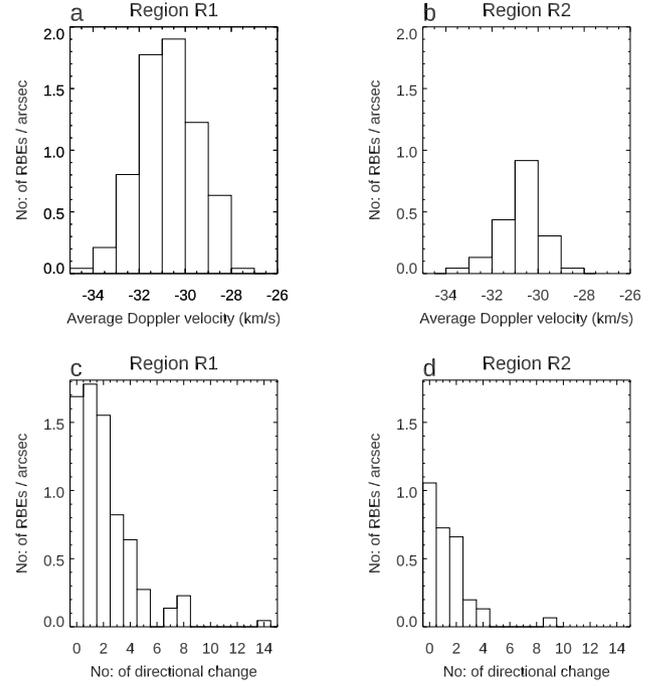}
\caption{Panel (a): Distribution of average Doppler velocity in region R1. Panel (b): Distribution of average Doppler velocity in region R2. The RBEs reverse its direction of horizontal motion (swaying) multiple times during its lifetime. The distribution of the number of direction reversals for the RBEs detected in R1 and R2 is shown in panel (c) \& panel (d), respectively. } 
\label{fig66}
\end{figure}

\begin{table}
  \caption{RBEs statistics for R1 And R2}

\begin{tabular}{|p{3.9cm}||p{1.5cm}|p{1.5cm}|}
 \hline
 \multicolumn{3}{|c|}{Region} \\
 \hline
 Properties     &R1&R2\\
 \hline
 Number/arcsec                     & $~7.11$          & $~2.81$   \\
 Lifetime (s)                      & $~82~\pm~59$      & $~64~\pm~36$   \\
 Transverse amplitude (km/s)   &$~12.3~\pm~4.3$   & $~12.3~\pm~4.0$  \\
 Doppler velocity (km/s)                   &$-30.7~\pm~1.3$   & $-30.7~\pm~1.0$ \\
 No: of direction change.          &~2                &~1\\
 Integrated transverse displacement (km)               &$~435\pm~316$          &$~345\pm~201$   \\
 Maximum transverse displacement (km)                     &$~197\pm~108$          &$197\pm~96$\\
 \hline
\end{tabular}
\end{table}

The length of the blue dots in the time-distance plot represents the lifetime of the RBEs. These blue dots are not strictly vertical which implies that the RBEs are undergoing horizontal motion. Based on this, we obtained the lifetime for each RBE and carried out a comparison between regions R1 and R2. The results are shown in Figure \ref{fig55}. Panel (a) and (b) presents the lifetime histogram for regions R1 and R2 respectively. By comparing these two plots, it is clear that the lifetime of RBEs in region R1 and R2 is similar. However, there are very few long-lived events in region R1. Next, we investigated the maximum value of the transverse velocities of RBEs in region R1 and R2. 
 Histograms for the maximum transverse velocities of RBEs in region R1 and R2 is presented in panels (c) and (d) of Figure \ref{fig55} respectively. As shown in panels (c) and (d) there is no significant difference in the maximum value of the transverse velocities of RBEs in region R1 and R2. Finally, we calculated the average Doppler velocities of RBEs over their lifetime and compared them in region R1 and R2. The results are shown in Figure~\ref{fig66}. The results show that the average Doppler velocities of RBEs in region R1 and R2 are also similar with a peak Doppler velocity at $\sim$30 km s$^{-1}$.

  \begin{figure}
      \hspace*{-0.54cm}
\includegraphics[width=0.5\textwidth]{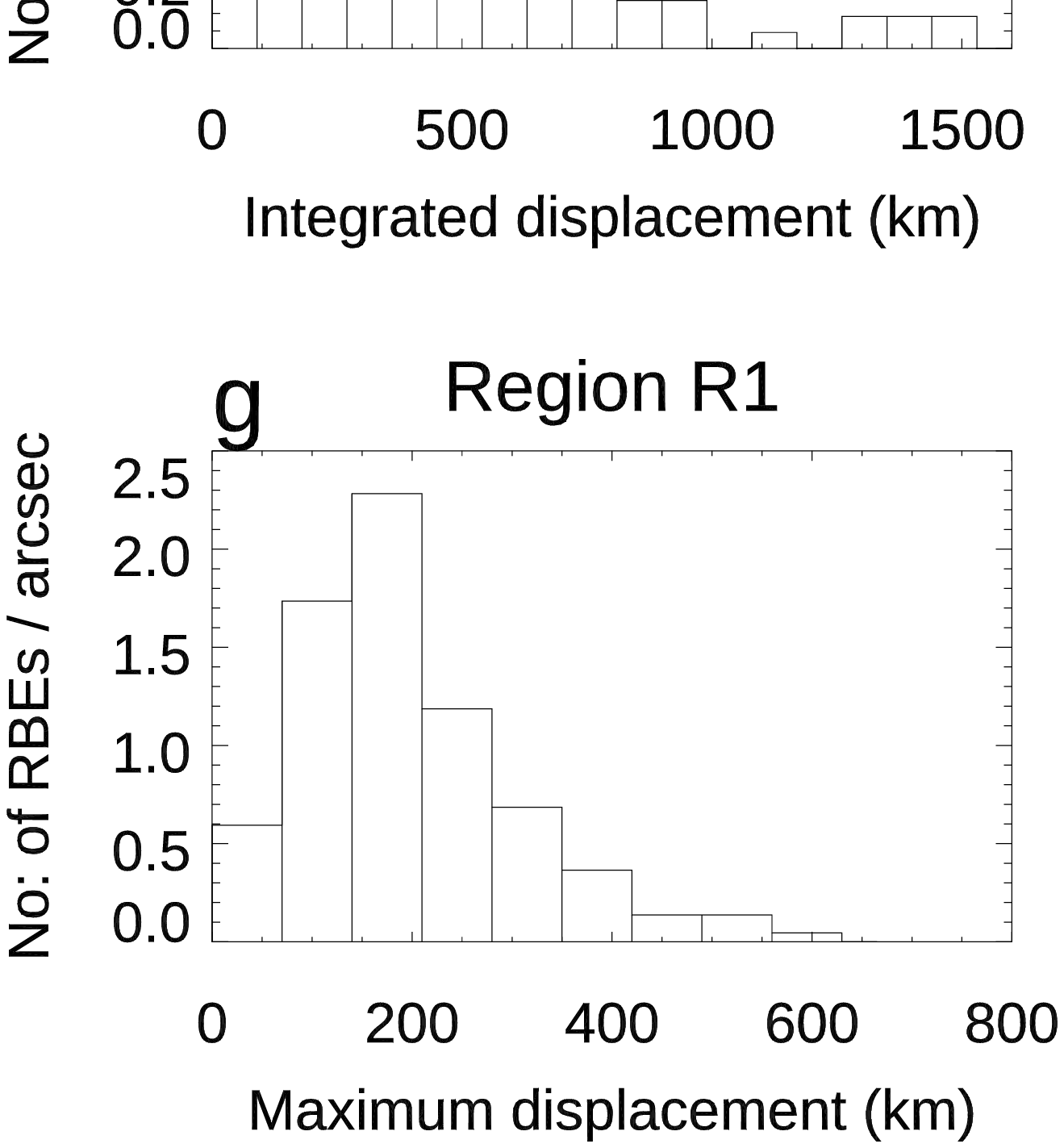}
\caption{ Panel (a): Lifetime of RBEs as a function of its number of sways in region R1. Panel (b): Lifetime of RBEs as a function of its number of sways in region R2. Panel (c): Integrated transverse displacement travelled by RBEs as a function of its number of sways for region R1. Panel (d): Integrated transverse displacement travelled by RBEs as a function of its number of sways for region R2. Panel (e) and (f) shows the histograms of the integrated transverse displacement of the RBEs in region R1 and R2 in one sway. Panel (g) and (h) shows the histograms of the maximum transverse displacement for the RBEs in region R1 and R2.} 
\label{fig77}
\end{figure}
\subsection{Swaying Motion}\label{swaying}

Next, we investigate the properties of the swaying motion of RBEs in the quiet Sun and the regions of the loop foot-points. The swaying motion can be identified from the time-distance plot by estimating the number of times an RBE reversed the direction of its horizontal motion. By manually inspecting the time-distance plot, we calculated the number of sways for 200 RBEs. No smoothing is performed when inspecting the time-distance track of RBEs. Moreover, RBE should move back-and-forth at least one pixel to be considered as a sway.  Histograms for the number of sways of the RBEs are plotted in panel (c) and (d) of Figure \ref{fig66}. From the histograms, it is evident that most of the RBEs in R1 and R2 sway 1--2 times during their lifetime. RBEs with zero sways, indicating the linear motion, are present in both R1 and R2 regions.   In Figure~\ref{fig77}, panel (a) and (b) we plot the lifetime of RBEs as a function of the number of sways it underwent. As expected, the number of sways increases with the lifetime of RBEs in region R1 and R2.  Next, we looked at the integrated transverse displacement travelled by the RBEs as a function of the number of sways. The results are presented in panel (c) and (d). The integrated transverse displacement is calculated by adding the absolute value of each transverse displacement underwent by an RBE during its lifetime}. In both regions, there is a clear linear relationship between the number of sways the RBEs underwent and the horizontal distance it travelled. 
Histograms of the integrated transverse displacement for the RBEs in region R1 and R2 are presented in panel (e) and (f) respectively. the distributions peaks at $\sim $ 300 km in both regions.
Finally, we also obtained the maximum transverse displacement, which is the distance between the two extreme transverse positions, and are presented in panel (g) and (h). Similar to \cite{sekse_2013_a}, the maximum transverse displacement obtained here is smaller than the integrated transverse displacement. This is expected since the RBEs are undergoing swaying motion.

\begin{figure}
\includegraphics[width=0.48\textwidth]{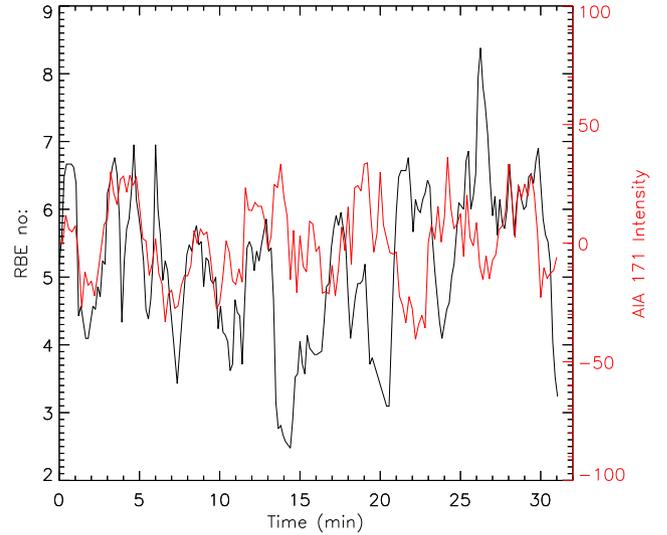}
\caption{The time evolution of the number of RBEs overplotted on detrended AIA 171 {\AA} (red) light curve.} 
\label{fig_corr}
\end{figure}

\subsection{Lifetimes, RBE motions}\label{liftetimes}
A study of the properties of the RBEs such as lifetime, Doppler velocity, horizontal motion, swaying motion reveals that they are indistinguishable between regions R1 and R2. Therefore, we confirm that the only parameter that differentiates regions R1 and R2 is the number density of RBEs. There is a larger number of RBEs in the loop foot-point region of the AIA 171 {\AA} channel image compared to the quiet Sun region. 

  \begin{figure*}
\hspace*{-1cm}

\includegraphics[width=1\textwidth]{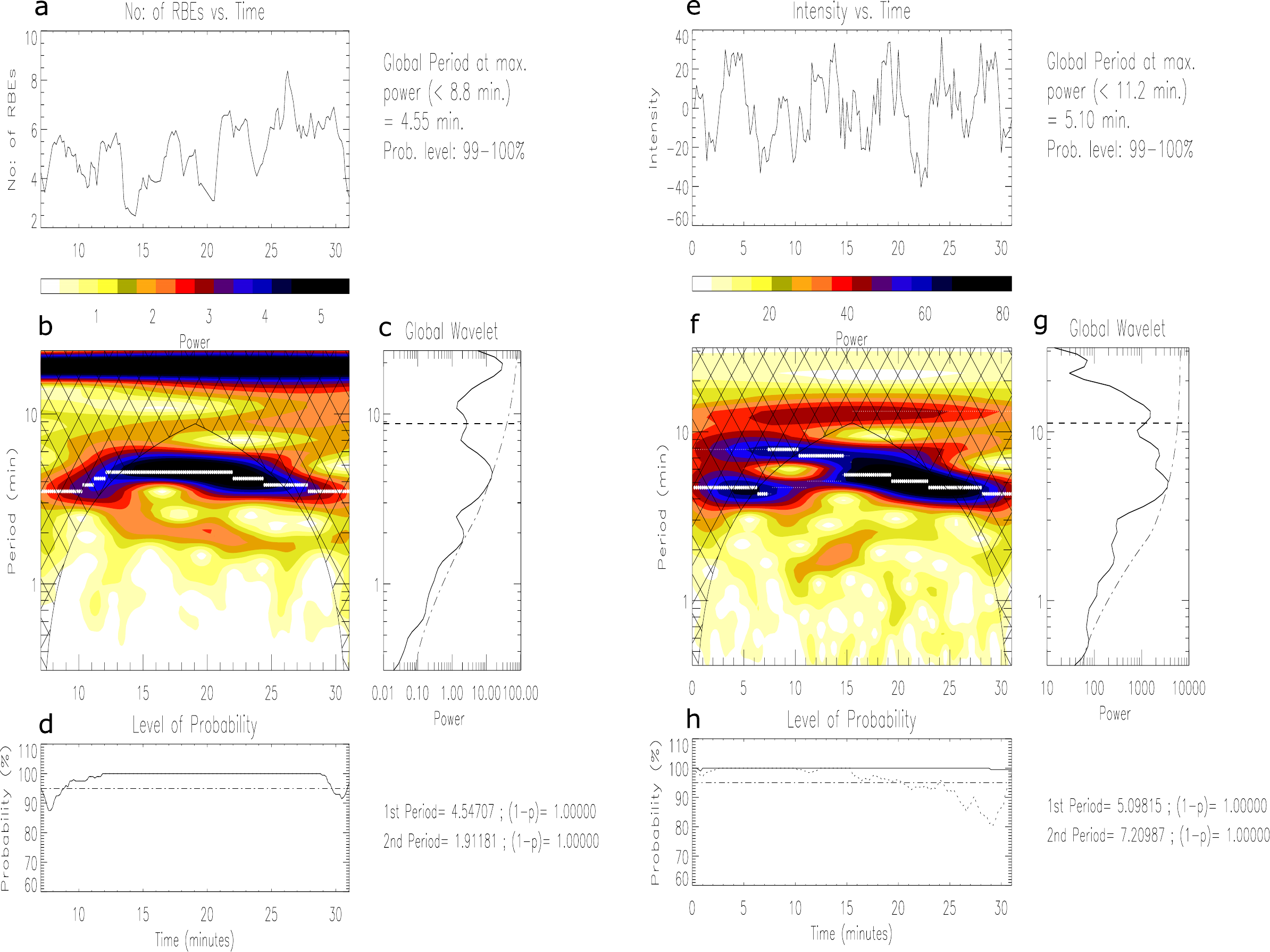}

\caption{left panel: panel (a) The time evolution of RBE numbers from 7 to 30 minutes, where the time starts at 10:58:54 UT on 27 June 2010. Panel (b): Local power spectrum. The thick dotted line marks the location of maximum power. Panel (c): Global power with the 95\% confidence level over-plotted.The peak at 4.55 min is above this significance level. Panel (d): The solid line shows the significance level obtained from randomisation while the dash-dotted line marks the 95\% significance level. Right panel; the equivalent plot for the AIA 171~\AA\ data over the time interval of 0 to 30 minutes.} 
\label{wlt_1}
\end{figure*}

  \begin{figure}

\includegraphics[width=0.5\textwidth]{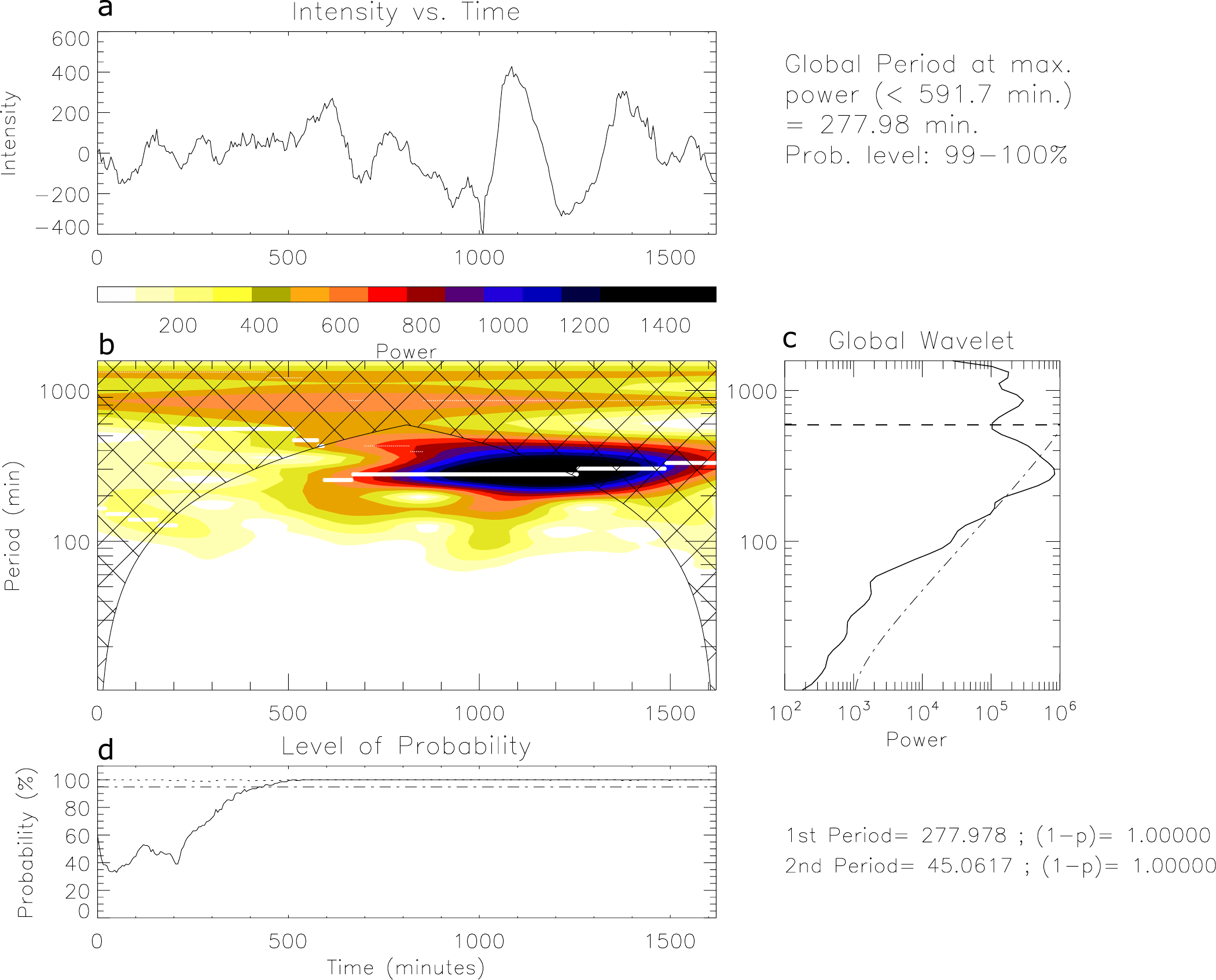}

\caption{left panel: panel (a) Detrended AIA 171 \AA\ intensity of the loop foot-point region over a $\sim$27 hrs interval. panel (b): Local power spectrum. Panel (c): Global power with the 95\% confidence level over-plotted. Right panel. Panel (d): The solid line shows the significance level obtained from randomisation while the dash-dotted line marks the 95\% significance level.} 
\label{wlt_16}
\end{figure}
\subsection{Heating \& periodic oscillations}\label{heating}

The role of RBEs in the heating of these coronal loops can be investigated by searching for heating signatures in AIA channels. We first detected RBEs in region R1 as explained in Section \ref{detection} and \ref{tracking}, using multiple slits at positions ranging from $y_{pix} = 120$ (5.1 Mm) to $y_{pix} = 140$ (6 Mm). 

We calculate the median, maximum, and average value of the number of RBEs observed in a region between 120 and 140 vertical pixels in the SST image to obtain the approximate number of RBEs. The time evolution of the median, maximum and average number of RBEs in the selected range shows similar behaviour.

To check the effect of activity variation of RBEs in the upper atmosphere, we then calculated the intensity averaged along a horizontal slit in AIA 171 and 193 {\AA} channel images in region R1. 
There is an increase in the RBE number density at a similar location as a local brightening in the AIA 171~\AA. 

During the reduced RBE activities, the light curve also shows a reduction in intensity. 

The AIA 171~\AA\ is more strongly correlated to RBE activities than AIA 193~\AA. Since there is a time delay in the AIA response, we performed the correlation analysis for two separate intervals. In the first interval between 0 to 14 minutes, the  Pearson correlation coefficient between the AIA 171~\AA\ light curve and RBE activity is 0.62, while for AIA 193 {\AA} the correlation coefficient is -0.09. In the second interval, between 14 to 30 minutes the Pearson correlation coefficient for AIA 171 and 193 {\AA} is 0.72, --0.48 respectively. In Figure \ref{fig_corr}, we compare the time evolution of RBE activity with a detrended AIA 171~\AA\ light curve. Until 14 minutes, there is a clear correlation between an enhancement in the RBE number density and the AIA 171~\AA\ intensity with a Pearson correlation coefficient of 0.64. Similar to the previous analysis, we see a time delay of 2 minutes in the detrended light curve for the interval between 14 to 32 minutes. This time delay is calculated by the shift in time which maximises the correlation coefficient between both curves. The correlation coefficient for the second interval is 0.72.

This may suggest that the energy input from RBEs is significant enough to heat the plasma to AIA 171~\AA\ temperatures, i.e. Log$_{10}$T=5.8. 
To check this better, we use a wavelet analysis.

\begin{figure}
\includegraphics[width=0.5\textwidth]{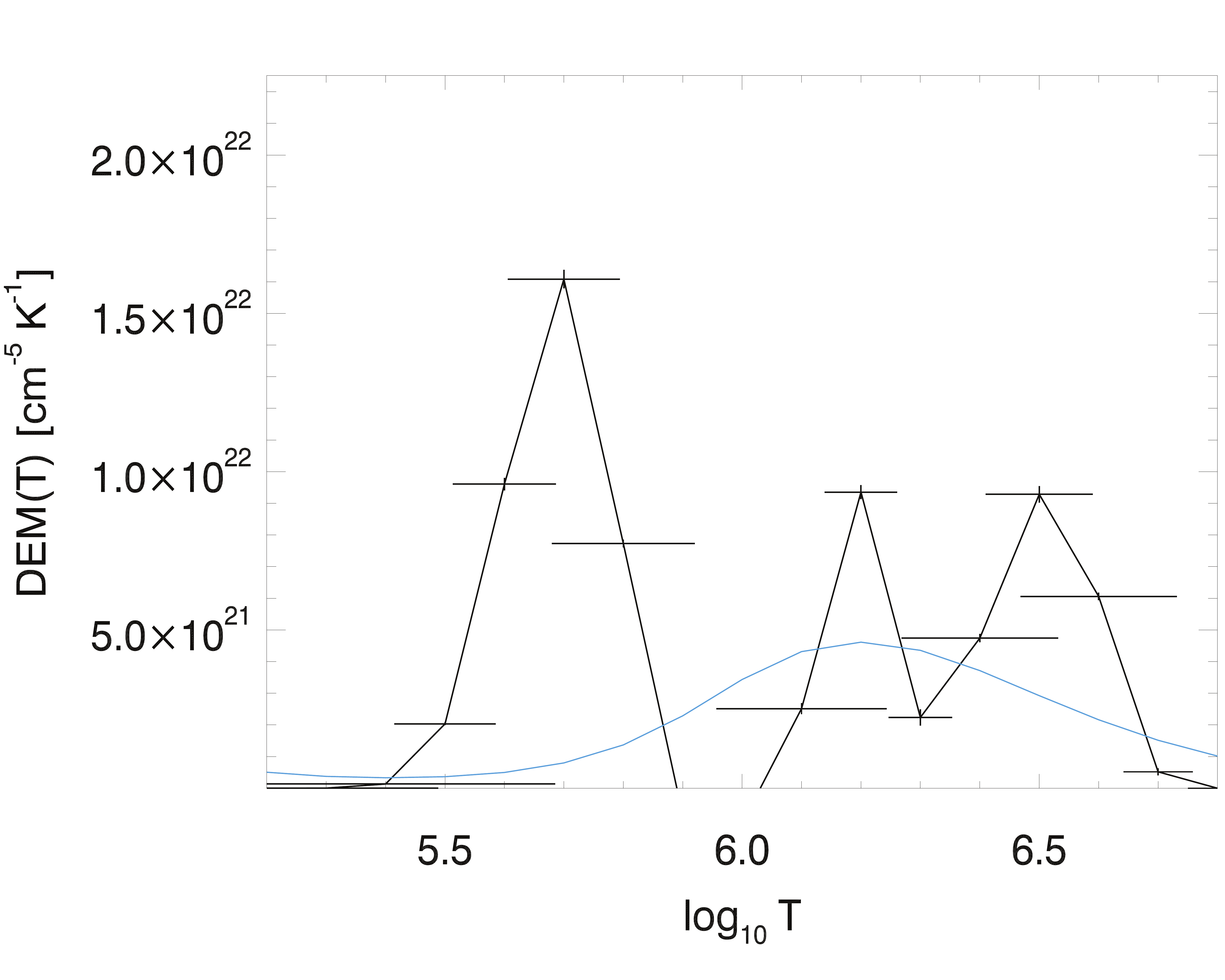}
\caption{SDO AIA regularised DEM for bright loop R1 section. The light-blue line represents the model spectrum.} 
\label{sdo_dem}
\end{figure}

 The detection of periodic oscillations and an evaluation of their significance is a challenging task. The analysis of periodic signals that we carried out in this work
represents a synergy of different methods: (i) a wavelet analysis \citep{TC} using the Morlet wavelet, with and without detrending  (ii) A Fourier transform method with confidence level estimated from the best fit to the power spectra.

It should be noted here that due to the intrinsic non-stationary properties of periodic oscillations (i.e. short lifetime, modulation of the oscillation amplitude and/or period) their oscillation energy often gets re-distributed across a number of Fourier harmonics, thus lowering the statistical significance of periodic oscillations in the Fourier analysis, e.g. \citet{Nak2019}.

The main advantage of the wavelet analysis over the Fourier transform is that it can provide the variation of the dominant period as a function of time. The results of the wavelet analysis for the average number of RBEs is presented in the left-hand panels of Figure~\ref{wlt_1}. The top panel shows the change in RBE number over the time interval of 7 to 30 minutes, where the data starts at 10:58:54 UT on 27 June 2010. This is where the periodic variation is strongest. The local and global power spectrum is shown in the bottom left, right panel respectively. The global power spectrum is calculated by averaging the local power over time. The cross-hatched region in the local power spectrum map represents the cone of influence (COI) where edge effects become important. The COI determines the maximum value of period that can be obtained from the wavelet analysis and it is shown as a horizontal line in the global power spectrum plot. For the AIA 171~\AA\ data, detrending was performed to bring out the local variation in intensity by subtracting the background from the original light curve, see the right-hand panels of Fig~\ref{wlt_1}. The background was calculated by smoothing the light curve over several time-steps, we selected ($\sim$480 s) which means that any detected period above this value was discarded. The background-subtracted light curve is shown in the top right panel. 

The wavelet analysis based reveals similar periodicities for the RBE number density/occurrence frequency and the corresponding variations in the AIA 171~\AA\ light curve with values of $\sim4.55$ min and $\sim5.10$ min as shown in Fig \ref{wlt_1}. The significance level for the RBE number density and the AIA 171 \AA\ flux is above 95\%. The bottom panels of Figure \ref{wlt_1} present the significance level obtained from the Fisher randomisation procedure \citep{Nemec_1985}. The advantage of the randomisation procedure is that the derived significance level is independent of the underlying noise model. The solid line in panel d \& h is given by $(1-p)\times 100 \%$. Here, the $p$-value is the probability that there is no periodicity in the data, which is given by the fraction of permutation that has a power greater than or equal to the peak power of the original time series. The derived significance value remains around the 100\% level for all of the time, therefore the periodicity can be considered as real.

The \cite{TC} code assumes red noise for the background noise model, which is a good approximation for a coronal time series with a power-law index of ${-2}$ \citep{Auchere_2016}. However, observations suggest that the power-law exponent varies between --1.72 and --4.95 \citep{Ireland_2015,Inglis_2015}. Therefore, the red noise model could lead to an erroneous estimation of the confidence level. To overcome this problem, we follow the code provided in \cite{Auchere_2016}, which estimate the 95\%  global confidence level based on the best fit to the Fourier power spectra. Here we fit the data using two different models as shown in equation \ref{eqn:eqn1} and \ref{eqn:eqn2}. 
\begin{equation}
P(\nu)=A\nu^{s}+C
\label{eqn:eqn1}
\end{equation}
\begin{equation}
P(\nu)=A\nu^{s}+BK(\nu,\rho,\kappa)+C
\label{eqn:eqn2}
\end{equation}
\noindent 
where $P$ is the power and $\nu$ is the frequency. The $K(\nu,\rho,\kappa)$ in equation \ref{eqn:eqn2} is the kappa function. In model 1 (eqn.\ref{eqn:eqn1}), we assume that the power spectra can be fitted with a simple power-law and a constant background (white noise). In the model 2 (eqn.\ref{eqn:eqn2}), the background power spectra is fitted with an additional kappa function. 

To obtain the periodicity in the activities of the RBEs, we performed the above wavelet analysis on the time evolution of the mean, median and maximum number of RBEs in the selected height range. We then performed a similar analysis on the time evolution of the AIA 171~\AA\ intensities to identify whether a  periodicity existed. A linear detrending was performed on the AIA 171~\AA\ data to bring out the local variations.

The wavelet analysis based on the \cite{Auchere_2016} code obtains similar periodicities in the RBE number density and in the corresponding AIA 171~\AA\ intensity. The periodicity obtained from the Fourier power spectra is the same for the time evolution of the mean, median and maximum number of RBEs: $\sim 4.45$ min and they are well above the 95\% confidence level. The periodicity derived from the time-averaged wavelet spectrum for the mean, median and maximum number of RBEs are 4.54, 4.54, 4.17 min respectively. For the detrended AIA 171 \AA\ light curve, the Fourier power spectra peaks at 5.20 min while the time-averaged wavelet spectra peaks at 5.09 min. The peak in the Fourier spectrum is slightly below the 95\% confidence level while the peak in the time-averaged wavelet spectrum is above the 95\% confidence level.

In addition, we have also searched the AIA 171~\AA\ data for evidence of long-term periods, Fig~\ref{wlt_16} shows evidence for a period of 4.6 hrs, we return to this in the Discussion section in relation to coronal condensation.

\subsection{Observed Regularised AIA DEM distribution}

\par To interpret the temperature contributions in the AIA FOV for each EUV channel in the coronal loop foot-point location where the spicules are sourced, we applied a Differential Emission Measure (DEM) using the generalised regularised inversion procedure, as developed by \citet{2012A&A...539A.146H}. The DEM ($\xi(T)$) quantifies the amount of plasma emitting within a certain temperature range and relates to the electron density of the plasma ($n_{e}$) as $\xi(T)\,=\,n_{e}^{2}\,dV/dT$, where $V$ is a volume element which will be determined from the pixel area of the emission measure (EM) maps and an estimate of the atmospheric column depth. 
\par \citet{2012A&A...539A.146H} constructed a model-independent regularization algorithm that makes use of general constraints on the overall form of the DEM {\it vs.} temperature distribution. Assuming optically thin emission in thermal equilibrium, the DEM is related to the observed dataset $g_{i}$ (observed intensity per AIA spectral channel ($i$) per pixel) as

\begin{equation}
\mathord{g_{i}} = 
\int_{T}\,\mathord{K_{i}}\xi(\mathord{T})dT + \delta\mathord{g_{i}},
\label{DEM}
\end{equation}

\noindent
where $\delta\mathord{g_{i}}$ is the associated error on $g_{i}$ and $K_{i}$ is the temperature response function (for AIA in this case). This ill-posed inversion problem needs to be stabilised. To do so the algorithm introduces extra information by way of a smoothness condition on the  AIA response function. By making a prior assumption of the smoothness factor and inverting the data with regularisation \citep{MR0162377}, the algorithm can reliably infer physically meaningful features, which are otherwise unrecoverable from other model-dependent approaches such as forward fitting (refer to \citet{2004SoPh..225..293K} for more information). This method has the added advantage of providing errors on the temperature bins through estimation of confidence levels when directly calculating the derivatives and then smoothing to a solution to return the DEM. The regularised inversion directly solves the minimisation problem, relating the data set to the expected CHIANTI \citep{1997A&AS..125..149D,1999A&AS..135..339L} DEM model. We refer to \citet{2012A&A...539A.146H} for a full description of the method which has been applied in many cases involving active region coronal loop systems observed with AIA. To derive the DEM, we consider the location of the brightest pixel in the AIA 171 {\AA} channel, which is located at the bright coronal loop section further along the loop leg above the path of the spicules. Then, we averaged the intensities of 9 pixels ($3\times 3$) centering on the brightest one.

It is clear from Figure~\ref{sdo_dem} that the DEM peaks at 1.6$\times$10$^{22}$~cm$^{-5}$~K$^{-1}$, at Log$_{10}$T=5.7 ($\sim 5.1 \times 10^5$~K), i.e. within the peak of the AIA~171~\AA\ pass-band. There is a smaller contribution from higher temperature channels, arising from the fact that there will be a multi-thermal line-of-sight contribution, within this FOV above the active region close to the limb. 

\section{Coronal loop simulations}

Simulations of coronal loop hydrodynamics were performed using ARGOS, a 1-D hydrodynamic code that solves the standard set of equations for mass, momentum, and plasma energy conservation for a fully ionized hydrogen plasma with a fully adaptive-grid package PARAMESH \citep{Antiochos_1999,MacNeice2000}. The code uses a fully adaptive-mesh-refinement (AMR) grid, that is, the spatial resolution of the grid cells varies to adequately resolve the computational domain where higher spatial resolution is needed. Specifically, the refinement/derefinement of the grid is performed based on the density gradient within a block of grid cells: if the density varies by at least 25\% in the block, then that block is refined (each cell of the block is split into two subcells); if the variation is less than 5\% then the block is derefined. This fully adaptive grid is necessary to adequately resolve thin regions of the computational domain where steep density and/or temperature gradients could possibly occur  \citep[see, e.g.,][]{Antiochos_1999}.

The code is based on the geometry of an arched loop with a total length $L = 200$~Mm and an apex height above the chromosphere of $h \approx 8$~Mm as described in \citet{Karpen2001} and \citet{Spadaro2003}. The loop foot-points include a 60~Mm thick chromosphere acting as a mass reservoir, with a temperature set to $T = 3 \times 10^4$~K. Hence, the full coronal section of the loop is initially $L_i=80$~Mm long, which is typical of active-region loops observed, for instance, by TRACE \citep[see, e.g.,][]{Aschwanden2000}.

The only boundary conditions imposed in the simulations are at the two endpoints (rigid wall, fixed temperature), which are located many gravitational scale heights deep in the chromosphere to minimize their effect on the simulated plasma evolution in the corona.
The temperature in the chromospheric part of the loop is kept constant throughout the run by forcing the radiative loss coefficient to vanish for $T\leq 29,500$~K and by limiting the density used in calculating the losses to $6\times 10^{11}$~cm$^{-3}$ \citep[see also][]{Spadaro2003}.

An initial equilibrium state must be achieved before applying additional impulsive heating to the foot-points of the loop. Failure to do this causes the loops to cool to a level where the loops become unsustainable.
This is done by supplying a spatially uniform and temporally constant background heating of $2\times 10^{-5}$~erg cm$^{-3}$ s$^{-1}$ to the whole loop. The chosen heating rate results in an equilibrium state with an apex temperature of about 750,000 K according to the \citet{RTV} scaling law at the end of the relaxation phase. The background heating is maintained in the chromospheric part of the loop throughout the simulation, but it is switched off in the corona when additional impulsive heating is supplied, 
 i.e. the heating rate is only applied in the chromospheric segments of the loop (0-65 Mm) and not in the loop segment which we simulate. Therefore, in the simulations, there is no background heating being applied in the coronal loop segment for any of the time stamps we report on in all of the runs.
The additional impulsive heating function is modelled using a Maxwellian temporal component and an exponentially decaying spatial component with a scale length of 10~Mm in the corona \citep[for further detail see][]{Susino2010}.

\begin{figure}
\includegraphics[width=0.5\textwidth]{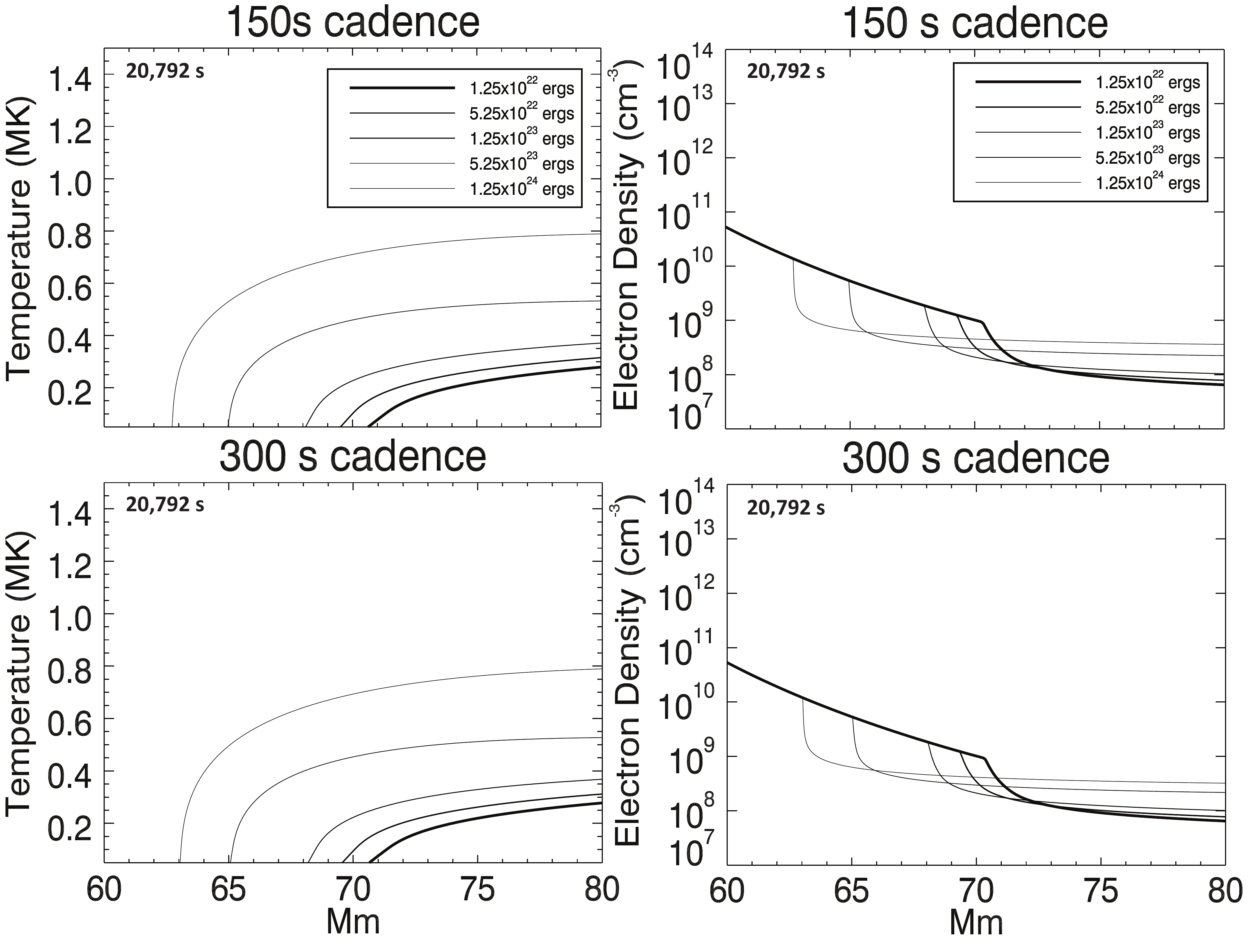}
\caption{Loop-leg properties from all simulations after loop relaxation (after 20,792~s), for 150~s cadence and 60~s bursts, plus 300~s cadence and 60~s bursts in all energy bands (thick line is lowest energy and thinner line is for higher energies)..} 
\label{fig2_sim}
\end{figure}

Since we take, by definition, the top of the chromosphere as the level at which the plasma drops below 30,000 K, the exact position of the top of the chromosphere ($s = \pm L_i/2$ from the top of the loop at the beginning of the simulation, $s$ being the curvilinear coordinate along the loop) changes during the calculation with the plasma filling or evacuating the loop. Hence, during the simulation, we find a new position for the top of the chromosphere $s = \pm L_f/2$ and, consequently, a new value of the height of the loop apex above the $T = 3 \times 10^4$~K level. A change in height of the chromosphere is clearly seen in Figure \ref{fig2_sim}. For the higher energy run, the top of the chromosphere is located at a height of $\sim 62$ Mm, while for the lower energy run the top of the chromosphere is at $\sim 70$ Mm.

The objective of this study is to simulate the coronal response due to different energy inputs close

to the loop foot-points, to investigate the evolution of the loop temperature and density, 
and in turn explore the modelled DEM for a hypothetical RBE to compare with the observations. We can increase the thermal energy contribution to the total energy input as a means to explore the expected loop response concurrent with the bright loop section (R1) compared with the dark loop section (R2), i.e. where we observe up to 10 times more RBEs per 60~s interval. For this, we have designed a series of simulations ranging from 10\% of an RBE thermal energy up to the energy equivalent to 10 RBEs where we take the thermal energy for one RBE as $1.2 \times 10^{23}$~erg, as derived earlier and in line with numerous previous studies.

After the loop has settled in an initial steady state with negligible plasma velocities (after $\sim 6.1 \times 10^{4}$ s from the start of the simulation), we turn on an impulsive heating consisting of a sequence of identical energy pulses injected into the loop foot-points at a constant cadence. The pulses have a fixed duration of 60 s, and the total energy provided per pulse ranges from $10^{22}$~erg to $10^{24}$~erg. To check the effects of pulse cadence on the evolution of the coronal loop, we considered two different pulse cadence times of 150~s and 300~s (see Table~2). Furthermore, to reflect the observed evidence that real coronal loops are not symmetric, we also considered an imbalance between the heating rates provided at the two foot-points of the loop, the heating at the right foot-point being 75\% of that at the left one. 
The energy range was chosen to explore the scenario, whereby only 10\% of the RBE thermal energy is converted into heating corresponding to $1.25 \times 10^{22}$ erg in a 60~s burst, and the scenario, whereby 10 times the RBE thermal energy is converted into heating corresponding to $1.25 \times 10^{24}$ erg.  Most of the lower energy runs settle into a quasi-static equilibrium state (see Figure~\ref{fig4_sim}, lower energy run). The temperature of the loop did not show any significant variation between $10^4$ to $2 \times 10^{4}$ s. For the runs which show condensations, the cycle repeats with a period of 5.2 hrs.

\begin{table}
 \caption{Simulation properties.}
\begin{tabular}{|p{0.4cm}|p{0.7cm}|p{1.2cm}|p{1.2cm}|p{1.0cm}|p{1.0cm}}

 \hline
Burst lifetime & Driver scale length      & Pulse cadence  & Energy per pulse  & Loop length  & Class\\
 (s)  & (Mm)        & (s)      & (ergs)  & (Mm)     &\\
 \hline
60 & $10$ & $150$ & $1.25 \times 10^{22}$ & $200$& Steady\\
60 & $10$ & $150$ & $5.25 \times 10^{22}$ & $200$& Steady\\
60 & $10$ & $150$ & $1.25 \times 10^{23}$ & $200$& Steady\\
60 & $10$ & $150$ & $5.25 \times 10^{23}$ & $200$& Steady\\
60 & $10$ & $150$ & $1.25 \times 10^{24}$ & $200$& Condensation\\
  \hline
60  & $10$ & $300$ & $1.25 \times 10^{22}$ & $200$& Steady\\
60  & $10$ & $300$ & $5.25 \times 10^{22}$ & $200$& Steady\\
60  & $10$ & $300$ & $1.25 \times 10^{23}$ & $200$& Steady\\
60  & $10$ & $300$ & $5.25 \times 10^{23}$ & $200$& Steady\\
60  & $10$ & $300$ & $1.25 \times 10^{24}$ & $200$ & Condensation\\
 \hline

\label{table2}
\end{tabular}
\end{table}

\begin{figure}
\includegraphics[width=0.5\textwidth]{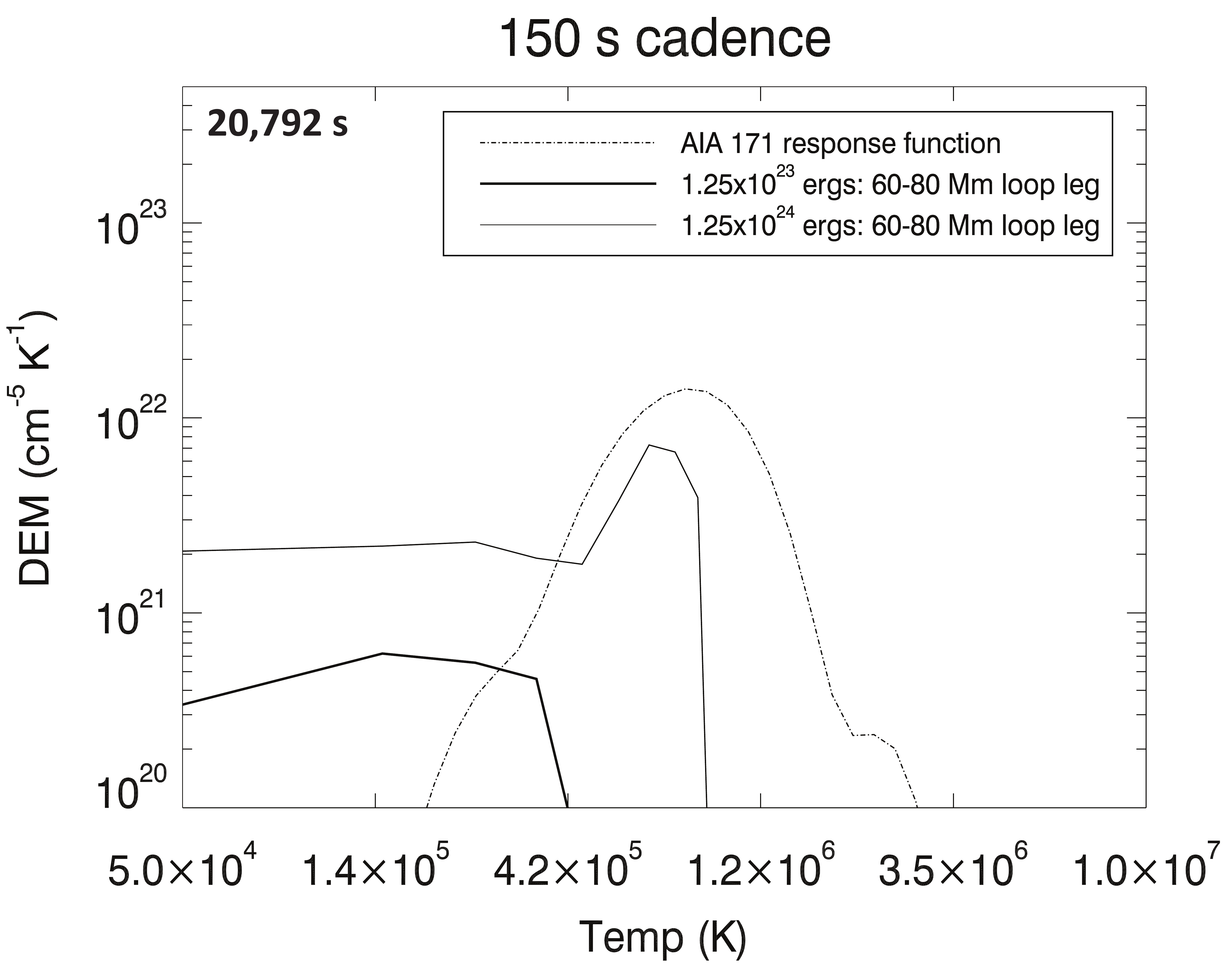}
\caption{{\it Solid lines}: Simulated DEM taken from simulated properties near the loop-leg after 20,792~s for two different energy input. {\it Dashed line}: SDO AIA response function for 171~\AA\ channel for comparison.} 
\label{fig3_sim}
\end{figure}
Figure~\ref{fig2_sim} shows the output of the simulations and provides the temperature and density variations along the loop for the 150~s and 300~s cadence runs. Firstly, we can report that the maximum temperature along the loop leg does not exceed 0.5 MK up to and including $1.2 \times 10^{23}$ erg, for both 150~s and 300~s simulations. One RBE energy input ($1.2\times 10^{23} $ergs) cannot account for the observed AIA 171~\AA\ emission signatures unless there is a large contribution into that pass-band from temperatures lower than 0.5 MK plasma (which could be possible given the broad pass-band of the filter). As shown by \citet{Vann2012}, the emission at this temperature comes in part from transition region lines plus a low temperature tail of Fe {\sc ix} 171~\AA. For the highest energy run in each series, the emission peaks at 0.8 MK in agreement with the AIA 171~\AA\ filter. Only the $5.2 \times 10^{23}$ erg  and $1.2 \times 10^{24}$ erg simulations (equivalent to there being between 4-9 RBEs per 60s burst), 
could possibly account for the strong emission signatures in the loop in R1.

In  Figure~\ref{fig3_sim}, we show the DEM following the integration of simulated loop densities over the full temperature structure corresponding to the 60-80~Mm loop-leg section after 20,792 s of temporal evolution. The DEMs are shown for two different energy inputs corresponding to $1.25 \times 10^{23}$ erg and $1.25 \times 10^{24}$ erg, in comparison with the response function of the AIA 171~\AA\ filter.  Note the DEM for the $1.25 \times 10^{23}$ erg simulation (in both 150 s and 300 s runs) is not sufficient for explaining the observed emission measures, although, can contribute to the pass-band of the AIA 171~\AA\ channel, with a marginal overlap due to plasma with temperatures in the range 2-4.2$\times$10$^{5}$~K. However, only the simulations with $1.25 \times 10^{24}$ erg (corresponding to $\sim$9-10 RBEs contributing to a burst) manage to reproduce a DEM peak at 0.8~MK on the order of 1$\times$10$^{22}$~cm$^{-5}$~K$^{-1}$, matching the observation very closely. 
The $5.25 \times 10^{23}$ erg runs (corresponding to $\sim$4-5 RBEs per burst) can achieve 0.5~MK, but the resulting DEMs peak at less than 10$^{22}$~cm$^{-5}$~K$^{-1}$.

\begin{figure*}
\includegraphics[width=1.\textwidth]{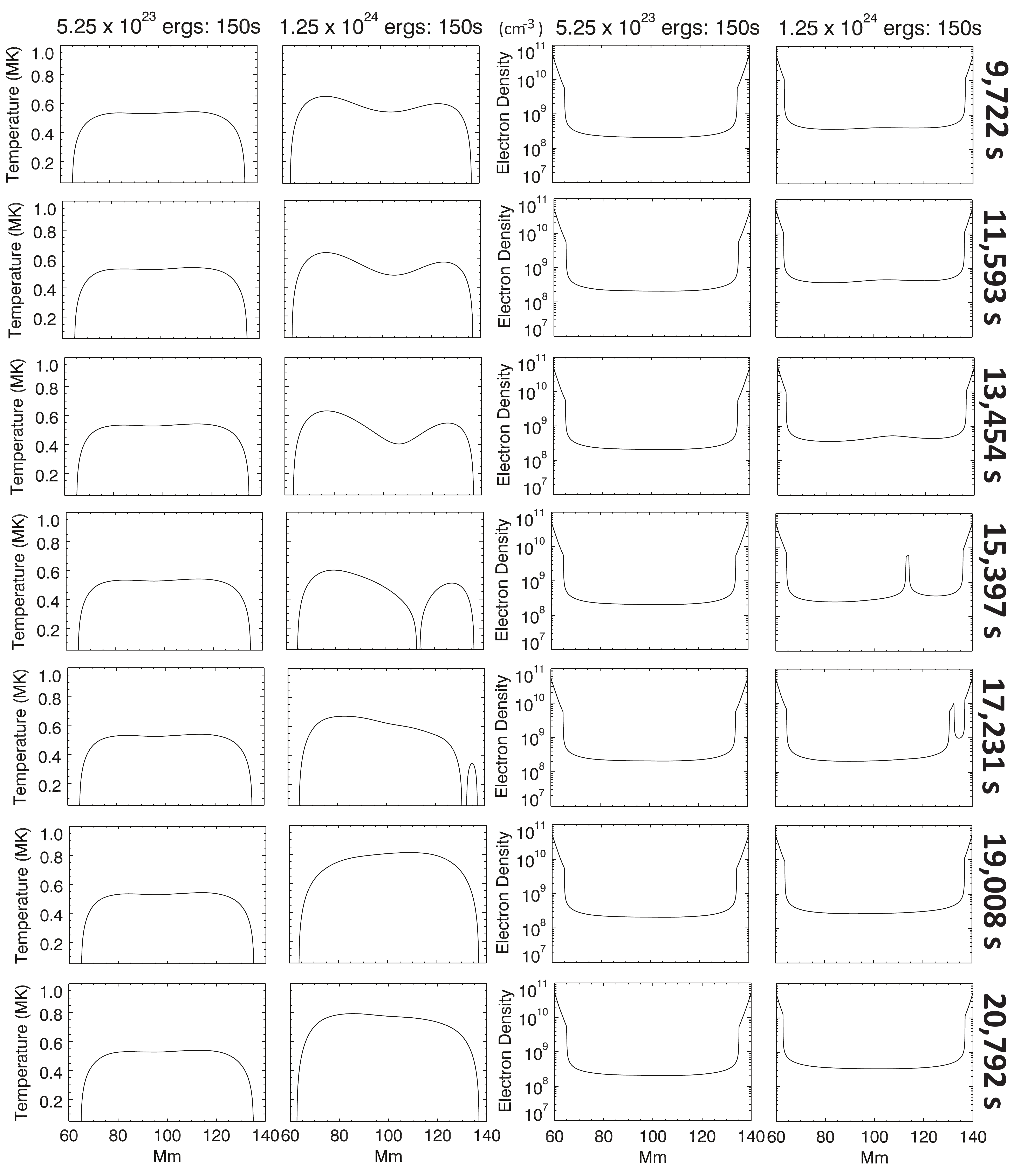}
\caption{{\it Column 1}: Temperature evolution of simulated loop for an energy input of $5.25 \times 10^{23} $ ergs per pulse. Over 5.27~hr from the start of the foot-point driver becoming steady. {\it Column 2}:  Temperature evolution for $1.25 \times 10^{24}$ over the same duration expressing signatures of catastrophic cooling near the loop-top and flow towards the foot-point. {\it Column 3}: Electron density evolution for $5.25 \times 10^{23}$ ergs/pulse, same as in Column 1, {\it Column 4}: Electron density evolution for $1.25 \times 10^{24}$ over the same duration expressing signatures of localised condensation near the loop-top followed by flow towards the foot-point.} 
\label{fig4_sim}
\end{figure*}

In Figure~\ref{fig4_sim} we show the electron temperature and density evolution over a 5.27 hr interval, revealing the onset of catastrophic cooling near the simulated loop-top only in the $1.25 \times 10^{24}$ erg run and not in the $5.25 \times 10^{23}$ erg run. Table 2 indicates which simulations evolve with a thermal instability establishing itself (labelled "Condensation") and those that do not undergo thermal instability (labelled "Steady"). The rapid cooling signatures followed by a localised density increase is followed by mass flow towards the opposing right-side loop foot-point, with respect to the left-side loop foot-point experiencing the higher energy input. Coronal loops are generally considered asymmetric in nature. Pressure difference induced by the asymmetric heating is responsible for the movement of cool blobs towards the less heated foot-points. In conclusion, only the $1.25 \times 10^{24}$ erg simulations (in both 150s and 300s runs) presented here provide sufficient coronal loop temperature and density structures that compare closely with the observed loops. Furthermore, only the $1.25 \times 10^{24}$~erg simulations result in loops that achieve the criteria for thermal instability, leading to mass condensations and mass drainage, thereby completing the mass-cycling between the chromosphere and corona. Due to the inability of the loop to attain a thermal equilibrium under the specific foot-point heating conditions, there is a thermal non-equilibrium. We see a timescale of 5.27 hr in the simulations from the onset of foot-point energy input to the drainage back to the opposing foot-point, which underwent a further 3 cycles during the full time range of the simulation.

\begin{figure}
\includegraphics[width=0.5\textwidth]{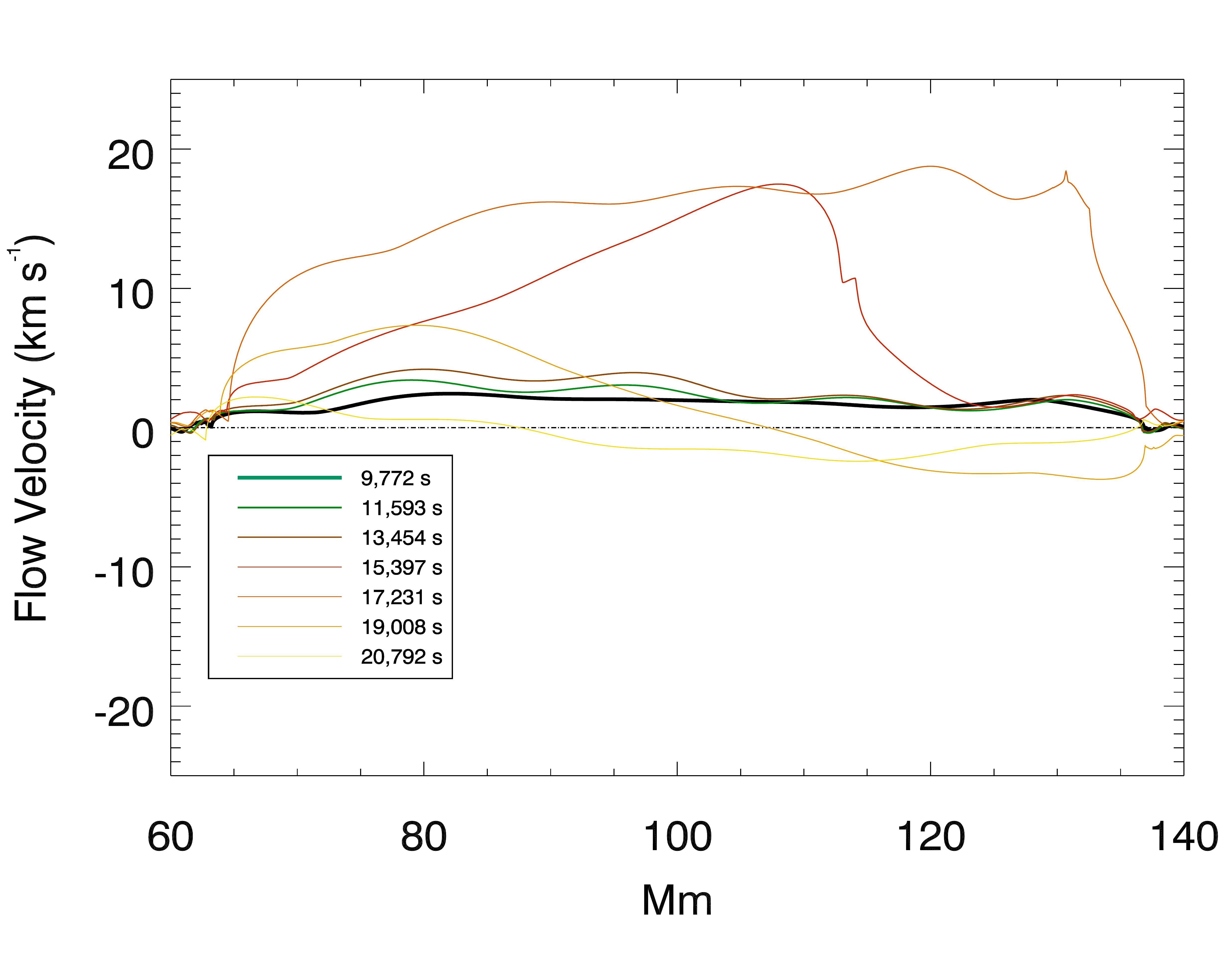}
\caption{The evolution of the mass flows along the loop over the 5.27 hr duration of the condensation event. }
\label{fig5_sim}
\end{figure}

In Figure~\ref{fig5_sim} we show the velocity evolution of the mass flows across the full length of the loop over the 5.27 hr duration of the condensation event. Throughout the mass cycle, there is a predominant net positive velocity flow (i.e. from the left foot-point to the right foot-point) which reaches a maximum of  19~km~s$^{-1}$, when the mass drainage occurs. The velocity flows imply motion in the direction of the coronal rain, i.e., from the loop top towards the less heated foot-point on the right, as can be seen
from Figure~\ref{fig4_sim}. However, relatively large positive flows $>$10~km~s$^{-1}$ appear throughout the loop after 17,231~s, which indicates that a global response of the loop during the condensation event (which otherwise appears to be highly localised near the loop-top when considering only the temperature and density changes). The strong flows which are present along the full length of the loop, result from the strong gradients of pressure that the loop responds to. As reported previously, e.g. \citet{Muller2003}, after the mass flows reach the rightmost foot-point we can also detect a rebound shock (with net negative velocities of 3-4~km~s$^{-1}$) between 19,008 s and 20,792 s that drives mass back towards the loop-top resulting in an oscillation of the entire loop structure. For an energy input of  $1.25 \times 10^{24}$ erg, we see a 5.27 hr periodicity in the occurrence of the coronal condensation; with a higher energy input the periodicity decreases, as reported in similar simulations, e.g. \citet{Susino2010}, with respect to hot loop conditions.

\section{Discussion}

We use high resolution observations of Type II spicules obtained with CRISP/SST and investigate their contribution to coronal loop foot-point heating. Type II spicules have been suggested to be a major supplier of mass and energy to the corona. After they fade with minor traces of down falling material (such as when observed in H$\alpha$), they continue to evolve in AIA (He {\sc ii} 304 {\AA}) and IRIS (Si {\sc iv} 1400~\AA, Mg {\sc ii} k) channels and falls back to the surface \citep{Pereira_2014,skogsrud_2015}. A rare case of Type-II spicule has been observed by \cite{Pereira_2014}, which shows falling material in Ca {\sc ii} H filtergram several minutes after the disappearance. These studies suggest that the rapid disappearance of the Type II spicules are indeed associated with heating, at least to transition region temperatures.

RBEs are the disk counterparts of Type II spicules \citep{Langangen}. In H$\alpha$ and Ca {\sc ii} 8542~\AA\ observations, they appear as a broadening in the blue wing with an unaffected red wing. Here, we detected such blueward asymmetries using the residual profile method following \cite{Rouppe_2009}. The lifetime of the detected RBEs ranges from 30 s to $\sim 200$ s. The lower cut off arises due to the minimum lifetime condition applied to avoid false detection. \cite{Pereira_2012} studied the Type II spicule in an active region using Hinode Ca {\sc ii} H filtergrams and found lifetimes to be 60 to 380 s with a mean value of $\sim150$ s. Using an extremely high cadence data (0.88 s), \cite{sekse_2013_a} found RBEs with a lifetime of 5 to 60 s in a quiet Sun region. Values of Doppler velocities obtained here matches with the range of velocities measured in previous studies of quiet Sun regions. Note that Doppler measurement is biased by the steep gradient in the line wing and cannot be regarded as solid line of sight velocity measurement. For RBEs studied in \cite{Sekse_2013}, the Doppler velocity distribution peaks at $\sim{26}$ km s$^{-1}$ while in \cite{Rouppe_2009} the peak is at $\sim{33}$ km s$^{-1}$. Transverse properties of the RBEs obtained here are also in agreement with previous studies. Similar to \cite{bart_2007}, the maximum transverse velocity peaks at $\sim$ 15 km s$^{-1}$. However, we do not see transverse velocities above 20 km s$^{-1}$. This is due to the constraint applied to differentiate between closely spaced RBEs. The integrated transverse displacement of RBEs detected here is larger than the maximum transverse displacement, which implies that the RBEs are undergoing swaying motion. The number of sways obtained by manual inspection reveal that the significant number of RBEs undergo 1-2 sways during their lifetime and it is slightly larger compared to \cite{sekse_2013_a}.

Although we do not see any difference in the spectral properties of Type II in a quiet Sun region and in foot-points of hot coronal loops, we find a significant difference, as much as a factor of 10, in their number density/occurrence frequency.  Looking at the time evolution via a wavelet analysis we detect a 5 min periodicity in both the RBE number density and the AIA 171~\AA\ intensity. Our observational findings are compared with model simulations of coronal loops performed with a 1-D hydrodynamic code that solves the standard set of equations for mass, momentum, and plasma energy conservation.

The adopted code simulates the evolution of the loop in response to a localised impulsive heating at the loop foot-points, with the aim of reproducing the plasma parameters and DEM derived from AIA 171~\AA\ observations. The simulations with an energy input of $1.25 \times 10^{24}$ erg (corresponding to $\sim$9-10 RBEs contributing to a burst) manage to reproduce a DEM peak at 0.8MK on the order of 1$\times$10$^{22}$~cm$^{-5}$~K$^{-1}$, matching the observation very closely.

In one case - for $1.25 \times 10^{24}$ ergs energy input at the foot-point - the simulated loop undergoes thermal non-equilibrium (TNE) and thermal instability.
As shown by, e.g., \cite{Antiochos_1999}, \cite{Spadaro2003}, \cite{Susino2010}, the occurrence of thermal instability, known as catastrophic cooling, during the loop evolution and the formation of plasma condensations, is a consequence of the loss of energy balance between the heating provided at the loop foot-points and the radiative losses at the loop top.

Catastrophic cooling is triggered when the evaporation of chromospheric plasma caused by the energy supply at the foot-points is strong enough to cause a substantial increase of the density at the loop top. If the heating is steady or impulsive with a pulse cadence well below the characteristic radiative cooling time of the loop, the plasma injected at the loop top has no possibility to drain back: it becomes more and more cool as radiative losses become more efficient, and a condensation - i.e., a blob of cool plasma - forms. Eventually, the condensation falls towards one of the loop legs under the action of the gravitational force and the gas pressure gradient force. Note that the radiative losses in the solar transition region and corona depend strongly on the plasma density - they are indeed proportional to the square of the density - and slightly on the temperature - they peak around 0.1~MK and then decrease for higher temperatures, see e.g., \cite{Cook_1989}. TNE is a consequence of having a strong localisation of the heating towards the foot-points, combined with a high occurrence frequency of the heating events compared to the radiative cooling time, and the loop is forced to undergo cycles of heating (evaporation) and cooling (condensation), known as TNE cycles (or limit cycles). During the cooling stage of the TNE cycle, thermal instability can set in locally, leading to the generation of condensations, which appear as coronal rain \citep{Antolin2012}.  See \citet{Antolin_2020PPCF...62a4016A} and \citet{Klimchuk_2019SoPh..294..173K} for a discussion on the difference between TNE and thermal instability.

The simulations presented in this work have explored a range of energies (from $1.25 \times 10^{22}$ to $1.25 \times 10^{24}$~erg per pulse) that yield loop temperatures at the top between $\sim 0.3$ and 0.8~MK, and densities below $10^9$~cm$^{-3}$. In these conditions, the characteristic radiative cooling time evaluated using the formulation of \cite{Serio_1991} is always longer than 1700~s (for $T < 0.8$~MK), implying that the considered cadence times (150~s and 300~s) are always shorter than $\sim 1/5$ of the cooling time. This situation is similar to that explored in some simulations presented by \cite{Susino2010} (see, e.g., their run no: 9 in Table 1). The evolution of the loop with the higher energy input considered in this work ($1.25 \times 10^{24}$~erg per pulse, corresponding to 9-10~RBEs per 60~s burst) is nearly the same as that reported by those authors for similar energies ($10^{24}$~erg) and pulse cadence $\sim 1/4$ the estimated cooling time.

Impulsive heating localised at the loop foot-points with such cadences results in the formation of a thermal instability along the loop. The thermal instability gives rise to plasma condensation near the loop top, and subsequent drainage back to the foot-point, opposing that where the energy burst is released, due to the lower gas pressure at the non-heated foot-point. This cycle has a total duration of 5.27 hours for the simulation with pulse cadence of 150 s, in agreement with the observed period of 4.6 hrs present the AIA data, see Figure~\ref{wlt_16}. 

It should be noted that our numerical investigation is limited to 1D and does not include several factors that could strongly modify the TNE onset and its periodicity. For instance, the heating scale length, while justified to some extent by the spicule length, is a factor that is very hard to accurately estimate. A reduction of this parameter lead to a significant reduction of the TNE cycle \citep{Froment_2018ApJ...855...52F}. Similarly, the loop's cross-sectional area expansion and detailed temporal occurrence and energy release of the heating events may alter the TNE evolution \citep{Johnston_2019AA...625A.149J}. The match between the obtained TNE cycle periodicity and that observed should be therefore taken with caution. 

Note that for $1.25 \times 10^{24} $ergs, the average density achieved at the loop top just before the onset of the condensation cycle is a factor of $\sim 2$ lower than that achieved in the corresponding simulation reported by \cite{Susino2010} ($\sim 8 \times 10^8$ vs. $\sim 2 \times 10^9$~cm$^{-3}$). For lower energy inputs, the plasma density at the loop top is even lower - up to a factor of $\sim 5$ for the loop with the minimum energy input - and this most likely prevents the onset of thermal instability, even if the temporal characteristics of the heating regime do not change with respect to the simulation with the higher energy input. This yields in turn a “steady” evolution of the loop, with a slow fluctuation of both plasma temperature and density inside the loop.

This slow evolution is not able to reproduce the DEM derived for the AIA 171 \AA\ emission band. Only in the case of thermal instability and consequent plasma condensation and drainage is a satisfactory reproduction of the “observed” DEM is obtained. In this case, the DEM distribution derived from the simulation outcome is able to qualitatively fit the peak of the DEM reconstructed from AIA 171 \AA\ observations, in terms of magnitude and temperature location. This is an indication that Type II spicules associated with at least 4 RBE per 60~s burst could be responsible for the heating of coronal loops.

It is worth noting that, however, as pointed out by \cite{Susino2010}, the DEM is actually insensitive to the occurrence of thermal instability, because the sequence of plasma evaporation, condensation, and draining does not effectively redistribute the plasma over the temperature, condensation, and draining. The appearance of the DEM peak around $\log T \simeq 5.8$ for the simulation with an energy of $1.25 \times 10^{24}$~erg per pulse is a consequence of a heating regime with pulse cadence time shorter than the plasma cooling time and energy deposition sufficient to trigger evaporation of chromospheric plasma to nearly coronal temperatures observable in the 171~\AA\ band-pass of AIA.

On the other hand, the condensation-drainage-evaporation cycle which completes the mass cycle between the chromosphere and corona, can possibly account for the mass flows observed in coronal rain \citep{Martens_1983}.
The heated loop is expected to be detected first in the hotter EUV spectral lines, for example in AIA 131 and 211~\AA\ channels. Afterwards, when the loop becomes cool and denser, it may become visible in cooler lines such as those sampled in the AIA 171 \AA\ channel \citep{Viall_2016}. In the TNE scenario, the heating events are concentrated towards the loop foot-points. If sufficient energy is injected, chromospheric evaporation should follow. \cite{Sahin_2019} observed two warm coronal loops in AIA 171 \AA\ images. One of these loops was observed to be filled with plasma, after which the entire loop becomes visible in AIA 171 \AA\ before disappearing and reappearing again. This observation supports the energy and heat injection (via magnetic reconnection). \cite{Li_2020} also show bright emission seen in the loop foot-point in AIA 131 \AA\ and 304 Å images, which corresponds to cool coronal plasma emission from almost 1 MK as seen in AIA 171 \AA\ to 0.1 MK in AIA 304 \AA. In this case, the cold plasma condensations occur. These condensations flow towards the surface of the Sun along the loop legs, as coronal rain.

\section{Conclusions}

Using simultaneous ground and space based data along with 1D loop simulation we study the contribution of RBEs in heating coronal loops. The result of this work is summarised as follows.
\begin{itemize}
    \item Coronal regions dominated by hot EUV loops have a significant higher number density of RBEs compared to a nearby quiet Sun region.
    \item Properties of RBEs such as lifetime, transverse amplitude, Doppler velocity, integrated and maximum transverse velocity, number of directional changes are similar in both the quiet Sun and hot EUV loop region.
    \item The time evolution of the RBE activity is correlated with the AIA 171 {\AA} light curve. Furthermore, a wavelet analysis reveals that the time evolution of the number of RBEs matches a 5 min oscillation as seen in the AIA data.
    \item Using 1D hydrodynamic loop simulation with an energy input of $1.25\times 10^{24}$ ergs (corresponding to 10 RBE thermal energy) at the loop foot-point, we generate hot coronal emission matching the observed AIA 171 \AA\ emission.
    \item However, for simulation runs with reduced energy input, say $1.25 \times 10^{22} - 1.25 \times 10^{23}$ ergs, fails to reproduce the synthetic AIA 171 \AA\ signature.
    \item This indicates there is a loop mass loading requirement (due to a high rate of spicules occurrences at the foot-point of the loops) to account for the AIA 171 \AA\ signature according to both the observations and simulations reported here.
    \item The higher energy run of $1.25\times 10^{24}$ ergs also produces a condensation cycle with a periodicity of $\sim$ 5 hrs in agreement with a 4.6 hr periodicity observed in the AIA 171 \AA\ data. 
\end{itemize}

Our study suggest that the thermal energy associated with RBEs could be responsible for the heating of coronal loops.

\section*{Acknowledgments}
The authors are most grateful to the staff of the SST for their invaluable support with the observations. The Swedish 1-m Solar Telescope is operated on the island of La Palma by
the Institute for Solar Physics of the Royal Swedish Academy of Sciences in the Spanish Observatorio del Roque de los Muchachos of the Instituto de Astrofísica de Canarias. We thank Gregal Vissers for helping with CRISPEX software during the data analyses. We would like to thank the AIA team for providing the valuable data. AIA data are courtesy of the Solar Dynamics Observatory (NASA) and the AIA consortium. Armagh Observatory \& Planetarium and NVN studentship is core funded by the N. Ireland  Executive through the Dept for Communities. P.A. acknowledges support from STFC Ernest Rutherford Fellowship (No. ST/R004285/2). MM acknowledges support from STFC under grant No. ST/P000304/1 \& ST/T00021X/1. JGD would like to thank the Leverhulme Trust for a Emeritus Fellowship. 
The data underlying this article were provided by the solar physics group at RoCS (University of Oslo), we thank L.H.M. Rouppe van der Voort who was the PI of the original SST proposal.

\section*{Data availability}
The reduced data underlying this article will be shared on reasonable request to the corresponding author. However, potential users should first contact the solar physics group at RoCS (University of Oslo), and in particular L.H.M. Rouppe van der Voort who was the PI of the original SST proposal.

\bibliographystyle{mnras}
\bibliography{ref}

\end{document}